%% file: ms.tex
\documentclass[smallextended]{svjour3}       
\smartqed  
\usepackage{appendix}
\usepackage{amsmath}
\usepackage{graphicx}
\usepackage{lineno}
\usepackage{array}
\usepackage{longtable}
\usepackage{natbib}

\usepackage{makecell}
\usepackage{multirow}
\usepackage{multicol}
\usepackage{amssymb}
\usepackage[lined,boxed]{algorithm2e}
\usepackage{xcolor}
\usepackage{soul}
\newcolumntype{P}[1]{>{\centering\arraybackslash}p{#1}}
\newcommand*\patchAmsMathEnvironmentForLineno[1]{%
\expandafter\let\csname old#1\expandafter\endcsname\csname #1\endcsname
\expandafter\let\csname oldend#1\expandafter\endcsname\csname end#1\endcsname
\renewenvironment{#1}%
{\linenomath\csname old#1\endcsname}%
{\csname oldend#1\endcsname\endlinenomath}}%
\newcommand*\patchBothAmsMathEnvironmentsForLineno[1]{%
\patchAmsMathEnvironmentForLineno{#1}%
\patchAmsMathEnvironmentForLineno{#1*}}%
\AtBeginDocument{%
\patchBothAmsMathEnvironmentsForLineno{equation}%
\patchBothAmsMathEnvironmentsForLineno{align}%
\patchBothAmsMathEnvironmentsForLineno{flalign}%
\patchBothAmsMathEnvironmentsForLineno{alignat}%
\patchBothAmsMathEnvironmentsForLineno{gather}%
\patchBothAmsMathEnvironmentsForLineno{multline}%
}
\begin{document}
\title{Search and Rescue in a Maze-like Environment with Ant and Dijkstra Algorithms}
\titlerunning{SAR in Mazes with Ant and Dijkstra Algorithms}
\titlerunning{SAR in Mazes with Ant and Dijkstra Algorithms}

\author{Z. Husain   \and
        A. Al Zaabi \and
        H. Hildmann \and
        F. Saffre   \and
        D. Ruta     \and
        A. F. Isakovic}

\authorrunning{Z. Husain, et. al.}

\institute{Z. Husain \at
    Electrical and Computer Eng. Department -
    Khalifa University (KUST) -
    Abu Dhabi, U.A.E. \\
    \email{zainab.husain@ku.ac.ae}             \\
\and A. Al Zaabi  \at
    Electrical and Computer Eng. Department -
    Khalifa University (KUST) -
    Abu Dhabi, U.A.E. \\
    \email{amna.alzaabi@ku.ac.ae}        \\
\and H. Hildmann\at
    Intel. Aut. Sys. Group -
    Netherlands Organisation for Applied Scientific Research (TNO)\\
    \email{hanno.hildmann@tno.nl}
\and F. Saffre\at
    Technical Research Centre of Finland (VTT) -
    Espoo, FI \\
    \email{fabrice.saffre@vtt.fi}        \\
\and D. Ruta\at
    Etisalat - BT Innovation Center (EBTIC) -
    Khalifa University (KUST) -
    Abu Dhabi, U.A.E. \\
    \email{dymitr.ruta@ku.ac.ae}        \\
\and A. F. Isakovic\at
    Colgate University - Physics and Astronomy Dept. - Hamilton, NY, USA\\
    \email{iregx137@gmail.com / aisakovic@colgate.edu}        \\
\thanks{ICT Fund for Bio-Inspired Computing}
}
\date{Received: DD Month YEAR / Accepted: DD Month YEAR}
%
\maketitle
\begin{abstract}
With the growing reliability of modern Ad Hoc Networks, it is encouraging to analyze potential involvement of autonomous Ad Hoc agents in critical situations where human involvement could be perilous. One such critical scenario is the Search and Rescue effort in the event of a disaster where timely discovery and help deployment is of utmost importance. This paper demonstrates the applicability of a bio-inspired technique, namely Ant Algorithms (AA), in optimizing the search time for a near optimal path to a trapped victim, followed by the application of Dijkstra's algorithm in the rescue phase. The inherent exploratory nature of AA is put to use for a faster mapping and coverage of the unknown search space. Four different AA are implemented, with different effects of the pheromone in play. An inverted AA, with repulsive pheromones, was found to be the best fit for this particular application. After considerable exploration, upon discovery of the victim, the autonomous agents further facilitate the rescue process by forming a  relay network, using the already deployed resources. Hence, the paper discusses a detailed decision making model of the swarm, segmented into two primary phases, responsible for the search and rescue respectively. Different aspects of the performance of the agent swarm are analyzed, as a function of the spatial dimensions, the complexity of  the search space, the deployed search group size, and the signal permeability of the obstacles in the area.

\keywords{Search and Rescue, Ant Algorithms, Ant Colony Optimization, Maze exploration, UAVs, Drones, Civil Security, Public Safety, Smart City}
\end{abstract}
%
\section{Introduction}
Drones
are on their way to become a pervasive technology. They have been around for quite some time (The US military used drones as far back as the Vietnam War \citep{BROAD10071981}) but have recently become commercially available for civilian use and even hobbyists.
%
Todays UAVs are technological advanced devices capable of autonomous flight operations \citep{hildmann:drones3030059}, making the \textit{remote} piloting an option, not a requirement. Drones come in all kinds of shapes and forms as well as vastly different prices, depending on their intended use the required specifications \citep{ARC.report.2015}.

The use of drones as mobile and airborne sensing platforms in general \citep{hildmann:drones3030059}, and with a focus on disaster response and civil defense in specific \citep{hildmann:drones3030071}, has been discussed in the literature.
Especially for data collection purposes, autonomous UAVs are increasingly a viable alternative to using human labor, with key application areas being
precision agriculture, 
civil defense such as fire fighting, 
traffic management, 
to locate victims in the aftermath of a disaster. 

Many applications can benefit from the use of (semi-)automated aerial devices due to the reduced cost 
and the ease of access these devices have \citep{SwarmIsMoreThanSumOfDrones.2021}. UAVs are especially useful 
for civil defense and disaster response,
because they are (a) expendable, (b) fast moving and (c) capable of moving in 3D space. The latter is extremely beneficial when existing infrastructure has been wiped out 
as drones are quite capable of transporting and operating communication infrastructure, making them well suited to form dynamic communication and sensing networks \citep{PuigCadiz2018}. 

Natural disasters often result in significant loss of communication and data-collection infrastructure. The use of 
\textit{swarms} (formations of multiple UAVs that can to some extent operate as a single operational unit) 
is a topic that receives increasing attention. Be it
to act as mobile sensor networks, 
to monitor personnel or victims 
or for tracking and surveillance tasks in general, 
the benefits of being able to deploy such systems quickly is evident.
\section{Background}
We now provide some background for
Search and Rescue (SAR) operations (Section \ref{subsec:BG:SAR}) and
the use of wireless and mobile / ad-hoc networks in this context (Section \ref{subsec:BG:MobileWirelessNetworks}) and argue that drone based wireless networks have great potential for this application domain. Finally we briefly elaborate on the problem of finding the shortest path between certain nodes in a network and argue for the benefits of nature-inspired heuristics to do so (Section \ref{subsec:BG:ShortestPath}).
%
%
\subsection{Indoor Search and Rescue (SAR) operations for swarms of drones}\label{subsec:BG:SAR}
Until recently,  the majority of Search and Rescue (SAR) applications assumed, for practical reasons, autonomously operating \textit{ground}-based robots because -- for 2D environments -- solutions have been proposed for collision free movement of a team of robots \citep{10.5555/3398761.3399020} or the movement of a team while in formation \citep{10.5555/3398761.3398848}. This article targets drones 
but also simplifies the environment to be 2D. UAVs are proposed in the literature 
for 
e.g.,
the handling of materials that are either themselves dangerous for humans 
or are located in dangerous or inaccessible environments.
The flexibility 
of drones makes them well suited to serve as mobile / aerial urban sensing platforms 
for Search and Rescue operations or for aerial tracking, 
especially in inaccessible environments and for the detection of victims or victims' life signs.
%
%
\subsubsection{Swarm (Multi-agent) Search and Rescue (SAR) operations}
Of course the use of drones is not restricted to disasters, but as such events often result in partial or total loss of existing communication and data-collection infrastructure and cause substantial damage to buildings, drones have been deployed to provide
situational awareness (SA), 
deliver (medical) supplies 
and monitor human personnel in the field, 
e.g., for fire fighters. 
%
%
\begin{figure}[htpb]
    \centering
      \includegraphics[width=3.4cm, height = 3.4cm]{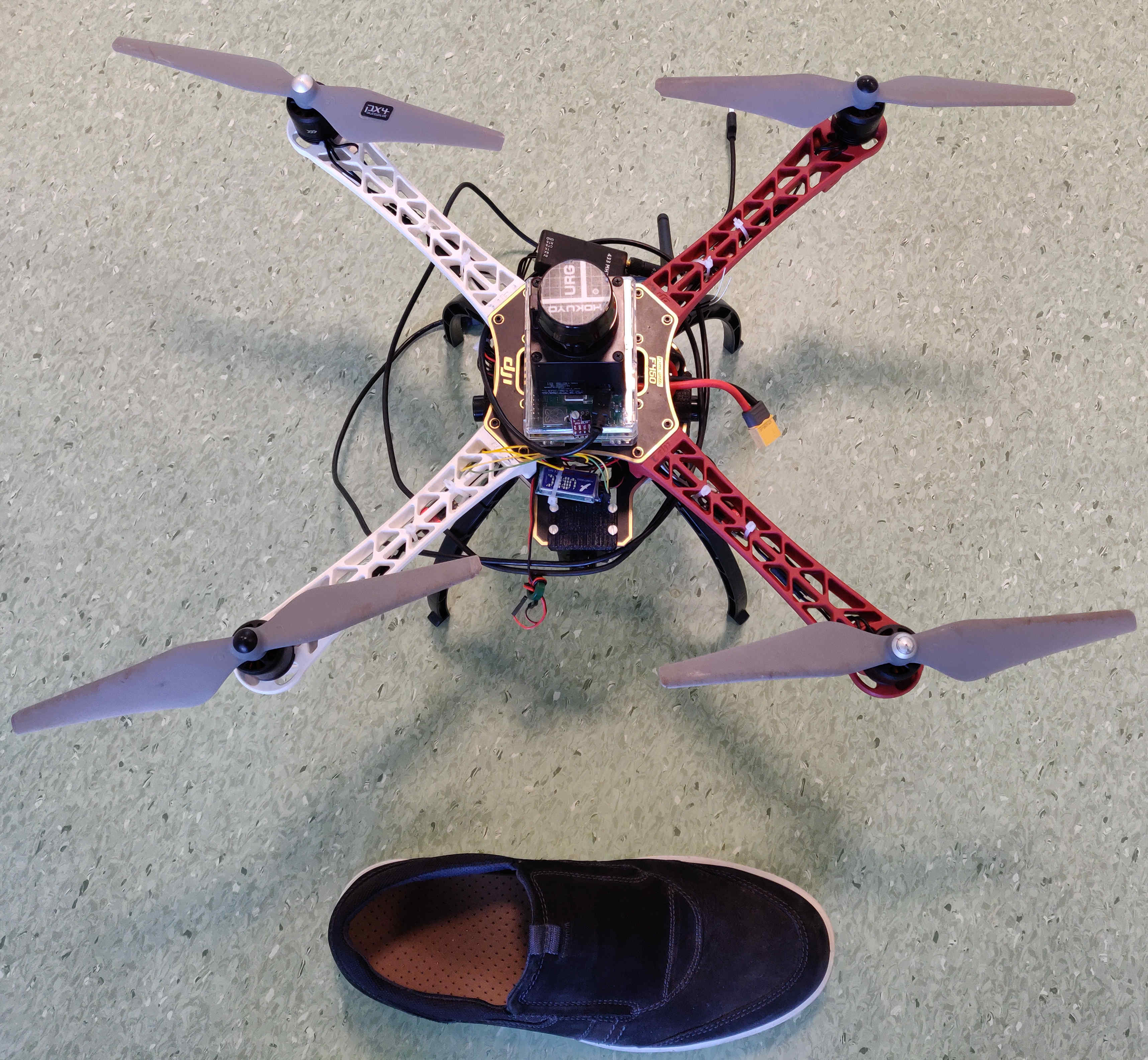}
      \includegraphics[width=3.6cm, height = 3.4cm]{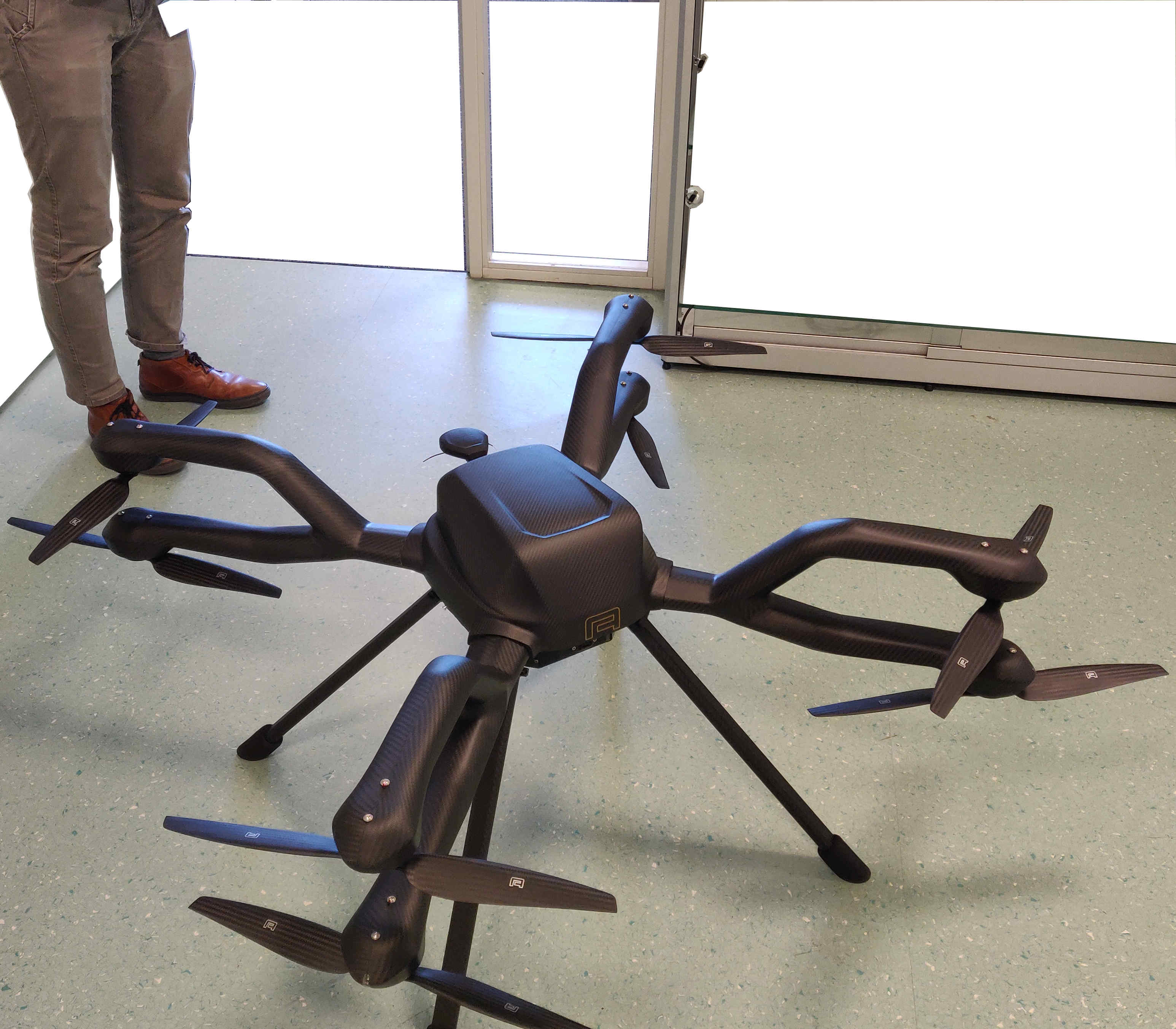}
      \includegraphics[width=4.2cm, height = 3.4cm]{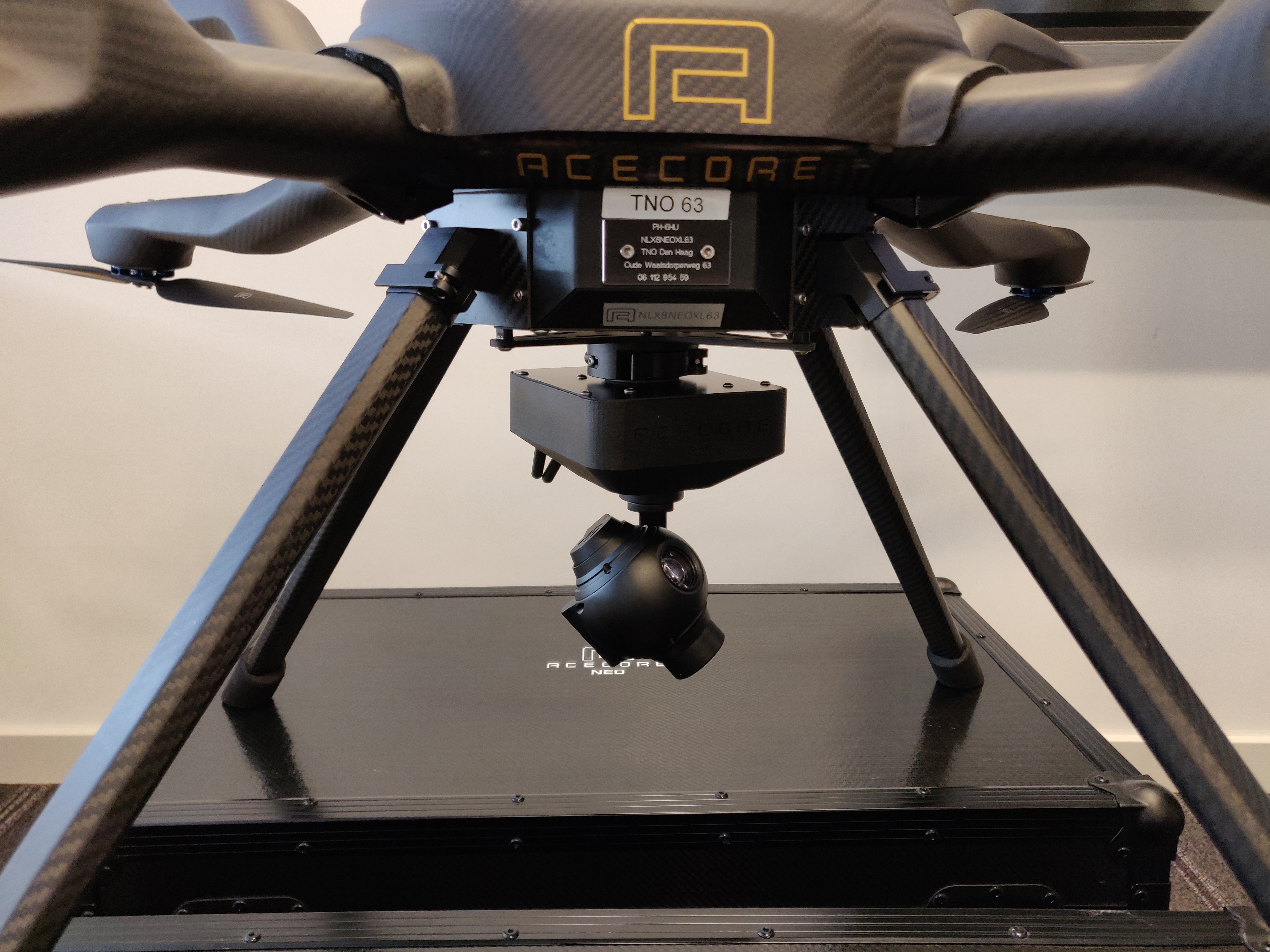}
\caption{Examples of the drones available at the Intelligent Autonomous Systems group at TNO: (left) a small device for proof of concept testing, shoe for scale. (center) a heavy-duty AceCore NEO drone. 
(right) a bottom mounted gimbal for visual data collection, possibly to be analyzed onboard for fast victim recognition.} 
\label{fig:hardware:TNO:AcrCORE}
\end{figure}

More recently, the drop in cost for units as well as the operation thereof has made the operation of multiple drones, acting as a single larger collective 
a realistic scenario for e.g., aerial tracking. 
\cite{Hussein:2014} have investigated SAR operations with ground based robot swarms using actual devices. The topic of performing SAR operations using autonomous agents has been an area of widespread research
and
several avenues are available for R\&D, including this work. There is still room for improvement in terms of optimizing the search algorithm for specific metrics such as the speeding up of the rescue. 

Considering temporal and spatial uncertainties 
in a disaster struck search space (subject to continuous changes),
heuristic techniques, rather than classic, deterministic ones are worth exploring in expectation of a better performance in such unstable environments \citep{a14030074}.
%
%
%
%
\subsubsection{Path planning in maze-like environments}
From an optimization point of view, the search in an ambiguous environment can be modeled as a maze solving problem where the maze walls denote the presence of unknown obstacles on the way to the rescue target.
Path planning through a prior known environment (\textit{``global path planning''}) is a well documented research topic, with several of its main issues having been addressed.
Maneuvering a maze with only local information of the search space is called \textit{``local path planning''}, which is performed \textit{on the go} during the maze solving process. Search in an unknown or uncertain environment is similar to a local path planning problem where heuristic techniques (cf. Section \ref{subsec:BG:ShortestPath}) can be employed to optimize the search process \citep{bonabeau1999swarm}. 
%
%
\subsubsection{Communication routing}
In the literature, communication services have been identified as an integral functionality of swarms 
and indeed, this can be of significant benefit while a relief team is deployed to reach the  victim at the rescue site.

This requires a routing algorithm, that can engineer a reverse route between the already dispersed search agents, under relevant communication constraints. There is a large number of such constraints and challenges for inter-UAV communication, we refer to \cite{7317490} for an overview over the literature.
%
%
\subsubsection{Summary}
\begin{figure}[h]
\includegraphics[width=\hsize]{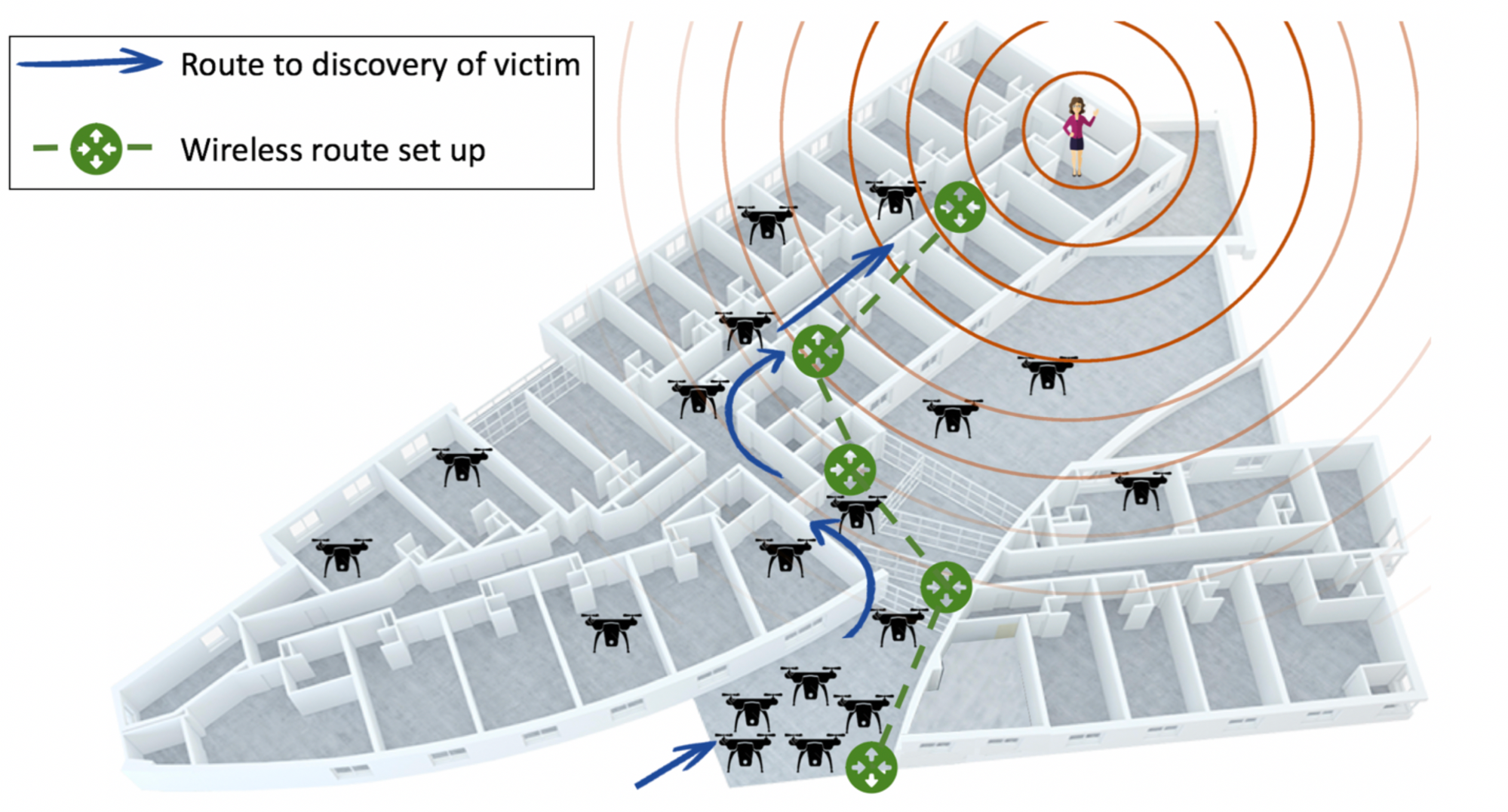}
\caption{An indoor Search and Rescue (SAR) scenario where the victim sends out a distress signal to guide the search by autonomous robotic agents. On discovery of the victim, the agents create an optimal relay link by optimizing various aspects like time, cost, length of the link, etc. to connect the victim to the base station. The other scattered agents not involved in the relay can be called back to the base.}
\label{fig:1}       
\end{figure}

The \textit{search} part of the SAR 
scenario is depicted in Figure \ref{fig:1}, where the agents maneuver a complex indoor setting to reach a target partially guiding the search using a beacon signal. Once the location of the target has been determined, the scattered agents realign to form the shortest possible relay link back to the starting point (base station). All unused agents return to the base, to prevent over-utilization of resources.
%
%
%
%
\subsection{Wireless Sensor Networks (WSN) in SAR for swarms of drones}\label{subsec:BG:MobileWirelessNetworks}
%
%
\subsubsection{Mobile Wireless Networks}
Mobile wireless networks 
have the potential to 
upgrade existing IT and communication infrastructure as, and where, needed. 
Communication networks have to be dimensioned to accommodate peak demands, which implies that outside those periods there is redundancy in the system. Given this redundancy, the ability to dynamically allocate additional capacity can dramatically increase the cost efficiency of the utilized resources. 
%
%
\subsubsection{Mobile Wireless Sensor Networks (MWSN) in SAR}
In case of an emergency situation, especially in SAR operations, fast, cost-efficient and simple deployment is a crucial factor when it comes to saving lives and reducing economic damage \citep{drones4030033}. Especially when the environment under considerations may have been altered or have suffered damage, the ability to deploy sensing capabilities alongside other equipment may facilitate real-time planning based on the gather information.
The more adaptive and reactive a deployment process into such an environment is, the more efficient as well as safer it can be realized.
Self-organization is increasingly applied to
this end
due to the fact that decentralized operations further reduce the need for communication infrastructure and speed up the decision making.
%
%
\subsubsection{UAVs as nodes in mobile networks}

The literature considers the use of UAVs as mobile nodes in communication networks or as parts of dynamic networks \cite{hildmann:drones3030059} to provide communication support 
and mobile communication infrastructure,
such as a drone-based (mobile) wireless sensor network (WSN) 
or wireless local area network (WLAN) in which communication is realized node-to-node (N2N) using ad-hoc 
routing 
for messages between nodes and generally, applications where the topology of the network is dynamic and adaptive. 
%
%
\subsubsection{UAVs in Search and Rescue operations}

Disaster response operations increasingly first deploy autonomous technology prior to dispatching human personnel. Drones can provide a first situational assessment and indeed, \cite{app9071474} highlights the ability to sense as an integral basic functionality of drones. Furthermore, the deployment of a UAVs (be it as single units or as a swarm) to act as a nodes in, or to entirely constitute, a cell network \citep{7133027} has been discussed in the literature.

For indoor operations, assets may have to operate in the absence of - or under the assumption of unreliable or compromised - communication infrastructures \citep{7165356}. Establishing a reliable means to communicate is often one of the first steps in deployment. Timely and efficient creation of infrastructure is often crucial but hampered by the uncertainty of the condition of the environment. The collision-free deployment process into an indoor environment \citep{11} can be complicated.
%
%
\subsection{Nature-inspired approaches to finding the shortest path}\label{subsec:BG:ShortestPath}
Cooperation and self-organizing collective behaviour \citep{selforganisation} has been studied to identify the underlying guiding principles, and has been successfully used them to increase the accuracy of models and to improve algorithms, cf. \cite{Navlakha2011}. One of the defining characteristics of approaches based on e.g. the foraging behaviour of social insects is their distributed nature. Distributed collective sensing has been shown to be possible using only rudimentary cognition and being cost-effective for computational agents \citep{Berdahl01022013}. Global goals can be achieved by locally processing the agents' (limited) view on the world. 
Computer scientists have studied models from theoretical biology and applied the underlying principles to computationally hard \citep{SLEEGERS2020100160} problems, often in the form of so-called \textit{heuristics} \citep{pearl1984heuristics}, which can find \textit{very good} solutions in extremely short time. For a compilation of nature-inspired approaches (including the algorithms) we suggest \cite{brownlee2011clever} (available online for free\footnote{http://www.cleveralgorithms.com/}).
%
%
\subsubsection{Deterministic versus heuristic search}
\input{Tables/table.litstudy.heuristics}
Deterministic techniques have been successfully applied to path planning \citep{mac2016heuristic, yang2011review}. Some prominent techniques include searches on Visibility Graphs (VG) and Voronoi diagrams (VD) \citep{leena2014survey}, Cell Decomposition Method \citep{mac2016heuristic}, and gradient techniques like the Artificial Potential Field Method \citep{Bounini17, khatib1986real, sutantyo2010multi}, amongst others.
Deterministic techniques, like the simple Wall Follower Algorithm, work optimally in maze search problems, provided the algorithm has complete knowledge of the search space beforehand \citep{hanafi2013wall}. When used in dynamic or unknown environments, deterministic techniques could end up in infinite loops or stuck in local optima, which could impede performance in SAR operations. Table \ref{table:litstudy.heuristics} summarizes the research in the field of maze solving using deterministic and heuristic approaches for both local and global path planning, which particularly impacted our thinking.
%
%
\subsubsection{The Dijkstra Algorithm}\label{subsubsec:BG:ShortestPath:Dijsktra}
Dijkstra's algorithm is a well-known search algorithm, developed by a Dutch computer scientist Edsger Dijkstra in 1959 \citep{nagatani2011redesign}, that can be applied on a graph. The algorithm is one of the solutions on Single-Source Shortest Path problem (SSSP) with directed or undirected graphs that has edges with non-negative weights \citep{ma2018path}. 
Dijkstra’s algorithm follows the greedy approach and finds the optimal path with the least total cost (shortest path from source to destination) \citep{xiaowei2016multi}.

The algorithm begins with a graph with nodes, \textit{u} or \textit{v}, weighted edges connecting nodes denoted as \textit{(u,v)} and weights as \textit{CostMatrix(u,v)}. The initiation of values and related steps before starting the path finding are:
\begin{itemize}
\item An array holding all edges costs (\textit{distance}) where all values are initiated to infinity except the first \textit{value(distance(source))} which is set to zero.
\item An array that contains all the nodes that have been visited during the search which by the end contains all nodes in the graph (denoted \textit{visited}).
Then, the algorithm proceeds as follows:
\item Until the array contains all nodes, we take the node with the least \textit{distance(v)} (starting with the source because \textit{distance(source) =} zero).
\item Node \textit{v} is then added to the visited array indicating that it has been visited.
\item Update distance values of adjacent nodes (\textit{u}) to the node \textit{v}.
\item If 
\textit{distance(v)} + \textit{CostMatrix(u,v)} \ensuremath{<} \textit{distance(u)}, 
then there is a new minimal distance founded for \textit{u}, so \textit{distance(u)} is updated with the new minimal value. Otherwise, no changes are made to \textit{distance(u)}.
Finally, after the algorithm visited all nodes in the graph and the smallest distance to each node is found, distance will now contain the shortest path tree from the source node \citep{mondada2002search}.
\end{itemize}

Dijkstra’s main advantage is its efficiency; its weakness are (i) the potentially large computational time and (ii) the inability to account for negative edges in the (search) graph. A brief, non-exhaustive summary of examples in the literature on using Dijkstra is shown in Table \ref{table:litstudy.dijkstra}.

\input{Tables/table.litstudy.dijkstra}
%
%
\subsubsection{Ant Colony Optimization (ACO)}\label{subsubsec:BG:ShortestPath:ACO}

In nature, several insects and animals have an inert sense of swarming together to accomplish a task more efficiently. Swarming, when used as a means for collaborating can help optimize both, the time and the energy spent.
Social insects such as ants effectively cooperate using \textit{stigmergy}, i.e., the ability to use the environment as shared memory / means of indirect communication to quickly determine a short path to some attractor.

The object of attraction in our case it is the search beacon, cf. Section \ref{subsubsec:SARinMaze:PhaseI.Explore:Approach}. Agents initially construct paths towards the beacon and deposit a \textit{pheromone} along their path. As time goes on, more direct paths have been used more frequently and are thus incentivized: the probability of an ant taking a particular path is represented by its density function given by
\begin{equation}\label{ACOeq}
P_A  (t+1)= \frac{(c+ n_A (t))^\alpha}{ (c+ n_A (t))^\alpha + ((c+ n_B (t))^\alpha}
\end{equation}
where \textit{c} is the degree of attraction to an unexplored path (the random element), \ensuremath{\alpha} is the bias to using a pheromone rich path, and \textit{A} and \textit{B} are the 2 paths to choose from at a fork . It is empirically demonstrated that \ensuremath{\alpha \approx 2}  and \textit{c} \ensuremath{\approx 20} often provide the best fit to experimentally observed behavior.

\section{Search and Rescue in Maze-like environments}
%
%
\subsection{Modelling the problem}

%
%
\subsubsection{Indoor search and rescue operations: a maze exploration problem}
By the very nature of the intended application domain, the environment for an indoor search-and-rescue operation is (at least partly) unknown.

Locating a specific position of a victim, reaching it and maintaining the path from that location to return to the entry point in such an environment is often, and aptly, modeled as a maze solving problem. The randomness of the maze walls can represent the random placement of obstacles in the area of interest. Therefore, the task of maze solving is essentially a path planning problem in an unknown environment. The main issues associated with maze solving include redundancy in found path \cite{Tjiharjadi6, Rivera12, Aurangzeb13}, and premature convergence while finding local optima \cite{Bounini17} \cite{Zhangqi11}. Several studies have attacked these problems using different path optimization techniques.

Therefore, we used a collection of mazes with different layouts, sizes and complexities. The mathematical model as well as a measure capturing the complexity of these mazes are presented in Section \ref{subsubsec:MaM:Models:Mazes}.
%
%
\subsubsection{The physical accessibility}
Mazes as a concept are rich with historical connotations, but abstractly speaking can be seen as environments with obstacles, commonly oriented such that elongated hallways or paths are created. In contrast to a labyrinth, mazes have branching points, that is, there are locations where the next step (which does not reverse the previous step, i.e., going back) can be in more than one direction. This gives rise to the most commonly known property of mazes, namely that the finding of a specific location or the retracing one's steps to return to the entry point, can be  challenging. While there are many variations on this theme, in the simplest form (and, as it turns out, a sufficiently general description) a maze is a collection of locations combined with a accessibility relation that determines, for any two locations, whether they are directly connected. Due to this, mazes can be represented as graphs \citep{diestel2017graph}.
%
%
\subsubsection{The signal accessibility}

While walls (obviously) block passage for humans entirely, their impact on signals is continuous, as radio signals can penetrate walls and other objects to a certain extent.
In the absence of blocking obstacles, a radio signal can be expected to extend outwards from the source homogeneously, with its strength decreasing with distance to the signal source or beacon.
\begin{figure}[h!]
\centering
\includegraphics[width=0.9\hsize]{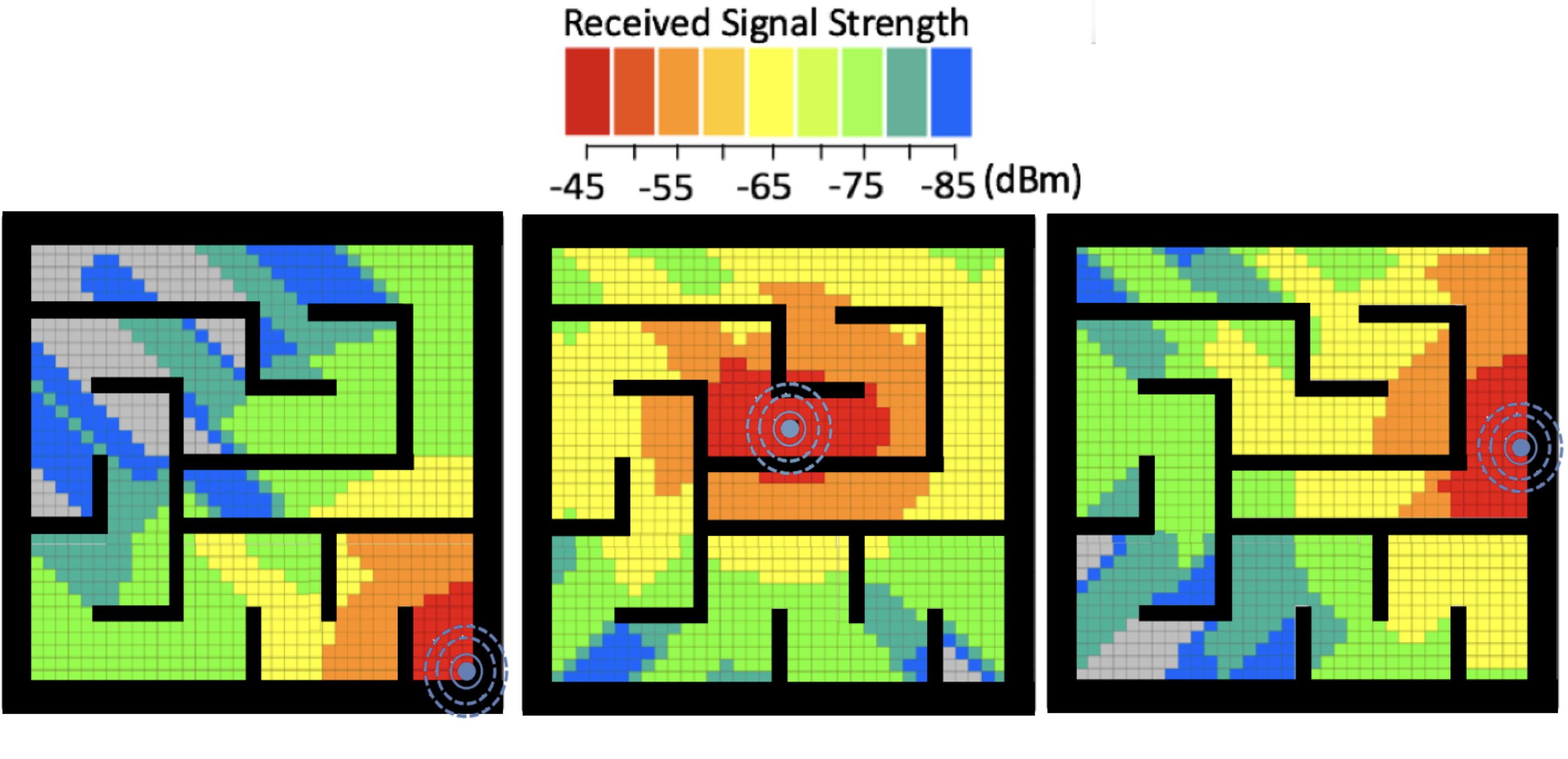}
\caption{Color-coded illustration of the beacon signal power received across the maze. The signal attenuates with distance and across the obstacles (walls) shown in black color. The attenuation is a function of the distance. The signal undergoes added attenuation across walls, due to the shadowing effect. }
\label{fig:5}
\end{figure}

Figure \ref{fig:5} (above) visualizes this for a specific scenario and beacon
in our scenario. The mathematical model for the signal strength and propagation is provided in Section \ref{subsubsec:MaM:Models:Routing} by Equation \ref{eq:4}.
%
%
\subsection{Search, and Rescue: a problem of two sequential phases}\label{subsec:SARinMaze:ExploreAndExtract}
\begin{figure}[h]
\centering
\includegraphics[width=0.8\hsize]{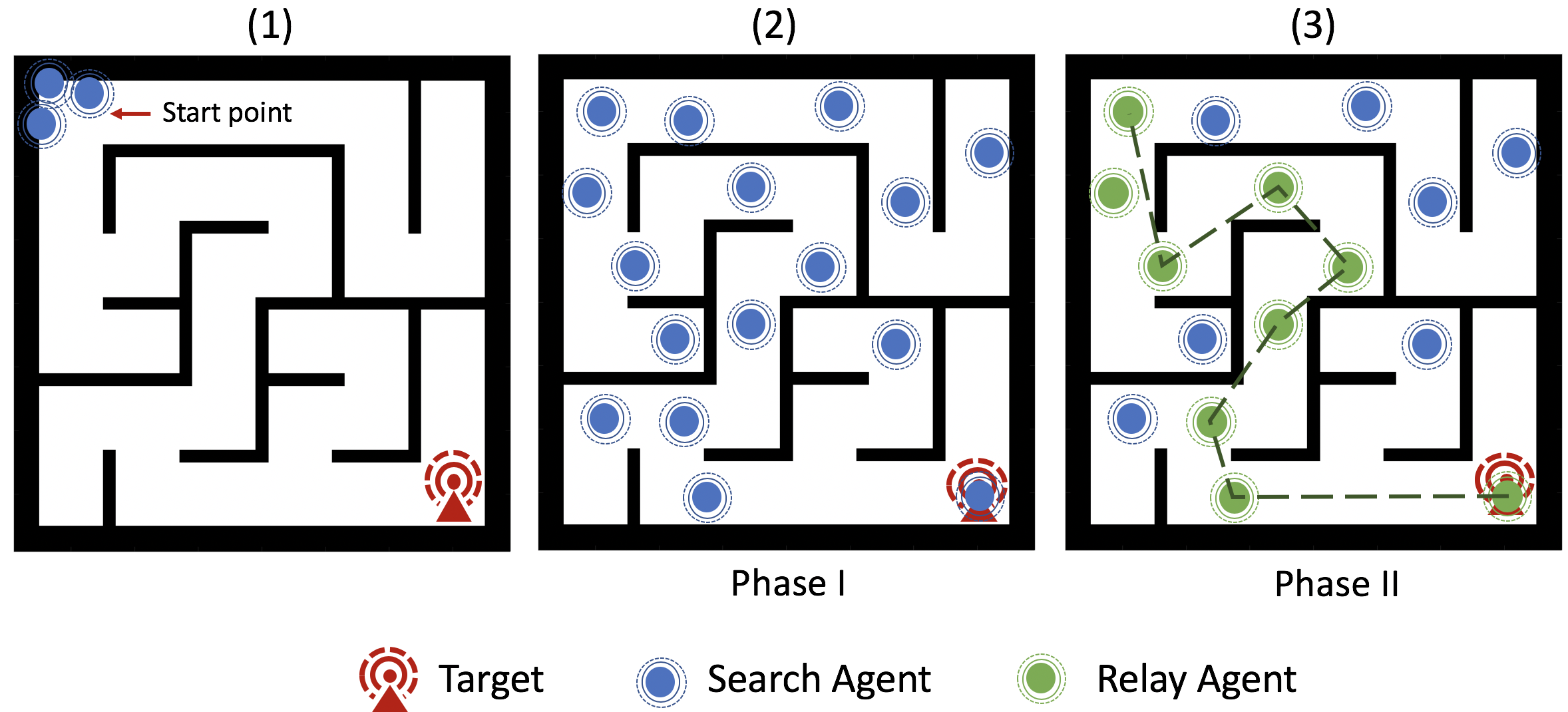}
\caption{The indoor search is translated into a maze solving problem.
Similarly to
Figure \ref{fig:1}, the agents traverse the maze in search of the target (depicting the victim) and then calculate the shortest and cheapest route amongst the scattered agents to form a relay network. }
\label{fig:2}       
\end{figure}

Figure \ref{fig:2} (on the next page) depicts the maze solving equivalent of the situation in Figure \ref{fig:1}, where the target is assumed to be at the ``end'', or a far corner, of the maze, and the agents are required to solve the maze and set up a reverse route, without any prior knowledge of its topology. This is the analogy that will be used throughout the course of this paper. Several tests are carried out by varying different aspects of the maze, such as maze dimensions and complexity, and the size of the search group.

%
%

\begin{figure}[h]
\centering
\includegraphics[width=0.9\hsize]{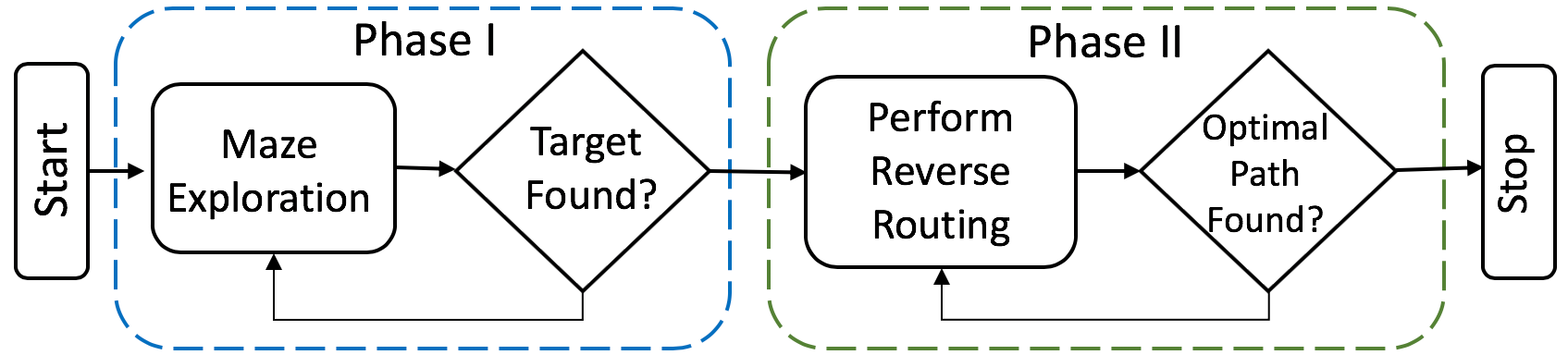}
\caption{The search and rescue process can be seen as two sequential phases: Phase I where the search is conducted in an environment; and Phase II where the extraction of an item or person from the environment is realized (i.e., the rescue part). For details see Section \ref{subsec:SARinMaze:ExploreAndExtract}.}
\label{fig:two.phases.SAR}
\end{figure}

Therefore, the SAR operation is translated into a two-phase process, where the search agents first collaboratively search the area to locate the victim, and upon discovery, shift to relay mode where a temporary relay network is set up using a subset of the originally scattered search agents, to complete the rescue operation. Figure \ref{fig:two.phases.SAR} shows the flow of control between the different sections of the algorithm through the search process.
%
%
%
%
\subsubsection{Phase I (Search): Maze exploration}
For the purpose of designing a path planning strategy for SAR, we prioritize finding a good path fast over finding the absolute shortest path. 
The path is further shortened in Phase II, during reversing routing in the Rescue Phase.
Our ultimate focus is on the search time, rather than path length. 

Therefore, in Phase I, a swarm of robots are injected into the search space. Each robots marks its explored territory in pheromone maps, local copies of which are shared between robots within a communication range. These pheromone map updates signal other robots to avoid already mapped/explored areas, quickening the overall exploration while shortening the search time.
%
%
\subsubsection{Phase II (Rescue): Signal routing and victim extraction}
Once the agents find the victim, they stop the exploration and the agent at the target finds a relaying route leading back to source node in order to pass the target’s information. This is done by finding and setting up the shortest possible link between the agents that are injected in the area. Among the possible methods to find the most optimal path, we employ a simple variant of the Dijkstra Algorithm  that uses the distance as the cost variable. As a route is calculated backwards to the source, the selected next-hop agents are kept fixed in position to form a relay network to assist the rescue operation.
%
%
\subsection{Solving Phase I (Search): Maze exploration}\label{subsec:SARinMaze:PhaseI.Explore}
The first phase of the Search and Rescue operation starts with the search agents rapidly exploring the maze, partially guided by the strength of the received beacon signal, as well as influenced by the movement of fellow agents.
%
%
\subsubsection{The Approach}\label{subsubsec:SARinMaze:PhaseI.Explore:Approach}
\begin{figure}[h!]
\centering
\includegraphics[width=0.8\hsize]{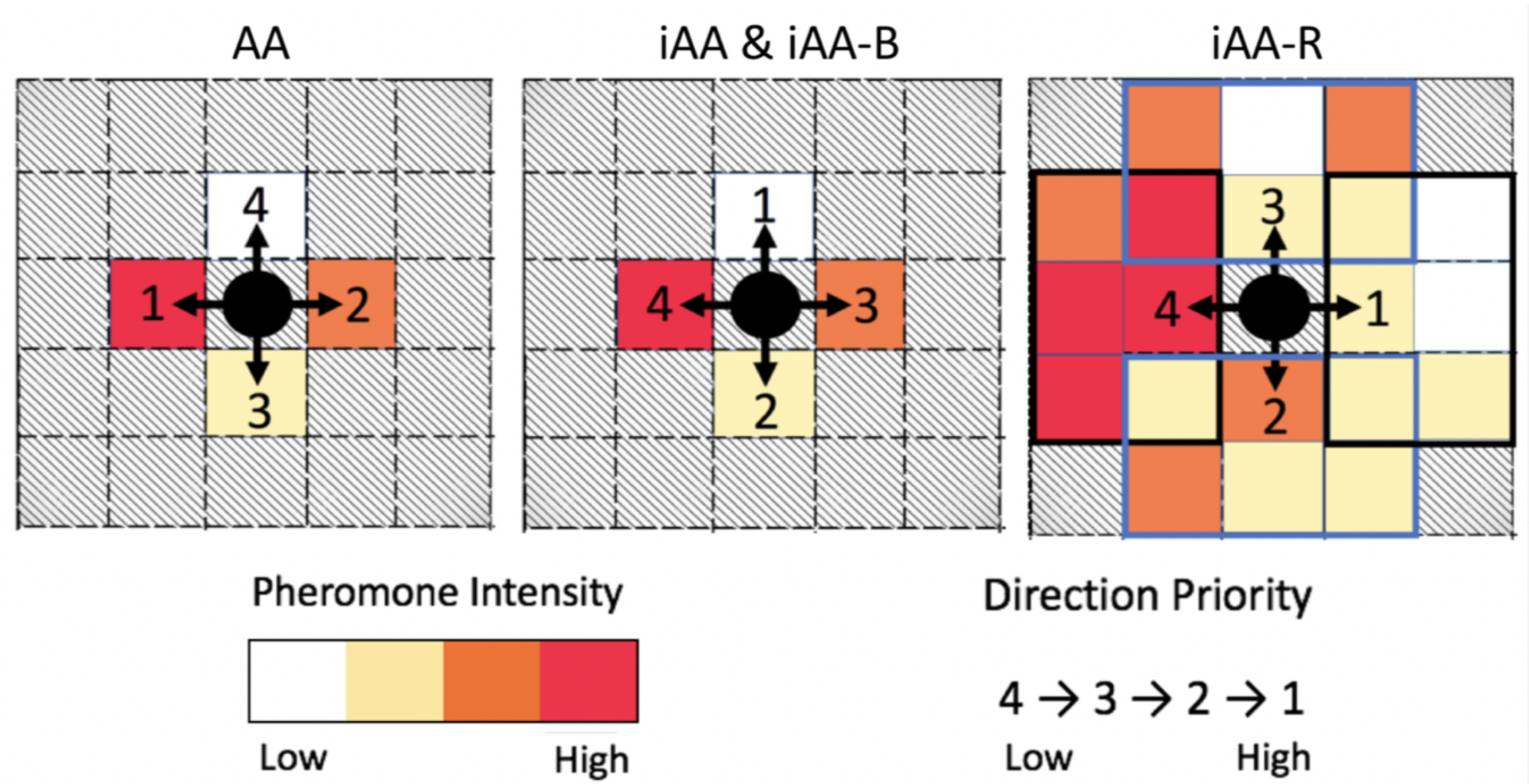}
\caption{All AA based decision making models prioritize their 5 possible directions of movement (1-4 are shown above, the option to remain stationary being the \ensuremath{5^{th}}).
AA prioritizes moving to a cell with a higher pheromone level, while all iAA based models prefer moving to a cell with lower pheromone concentration.
The iAA-R algorithm (rightmost image, above) checks for average pheromone in neighboring \textit{regions} while the other 3 approaches (AA, iAA and iAA-B) only check pheromone levels in the immediate neighboring \textit{cells}.}
\label{fig:6}
\end{figure}

A general procedure for the implementation of an ant system was introduced in Section \ref{subsubsec:BG:ShortestPath:ACO}. The AA local path planning based algorithm applies the ACO probability density function at each step to determine the next best step  using:
\begin{equation} \label{eq:5a}
P_i  (t+1)= \frac{(c+ n_i (t))^\alpha}{\sum_{j=1}^{5} (c+ n_j (t))^\alpha}
\end{equation}
where \ensuremath{P_{i}(t + 1)} is the probability of moving in direction \ensuremath{i}, \ensuremath{n_i(t)} is the amount of pheromone in block \ensuremath{i}. The summation (\ensuremath{j}) in Eq. \ref{eq:5a} is over the 5 possible directions of movement (cf. Figure \ref{fig:6}) with the \ensuremath{5^{th}} direction being remaining stationary.
Also, \ensuremath{c} is the degree of attraction to an unexplored path, and \ensuremath{\alpha} is the bias in using a pheromone concentrated path. An example of AA parameter values would be \ensuremath{c = 20}, and  \ensuremath{\alpha= 2}, as demonstrated in \cite{Engelbrecht11}.

With AA, however, ants were noticed to be clustering together, due to the attractive nature of pheromone which makes ants follow each other with little randomness. To counter this effect, and to quicken exploration, an inverted AA (iAA) model was developed with repulsive pheromones, encouraging ants to venture into unexplored territory. The probability density function used for decision making in iAA is
\begin{equation} \label{eq:5b}
P_i  (t+1)= \frac{(c+ n_i (t))^{-\alpha}}{\sum_{j=1}^{5} (c+ n_j (t))^{-\alpha}}
\end{equation}
With promising initial results, 2 new versions of iAA, namely inverted-AA with beacon initialization (iAA-B) and inverted-AA with an increased sensing range (iAA-R) were also developed. The iAA-R was developed to test the effect of a longer sensing range on the speed of convergence of the solution.

The different models differ only in terms of the Probability Density Function used to generate the roulette wheel distribution in the algorithm. Figure \ref{fig:6} visualizes the decision making based on the pheromone levels for all 4 models.
%
%
%
%
\subsubsection{The Algorithm}
The pheromone used to implement the AA based algorithms is virtual, stored as matrices in individual agents
and shared between agents.
The pheromone matrix helps the agents keep track of the pheromone intensities, while at the same time, helps them roughly map out the explored area.
\input{Algorithms/alg.iAA}

As each agent explores the search space individually, their pheromone matrices would differ. Hence, when 2 agents pass by each other, they exchange their copies of the matrix, to combine their search results, and come up with more inclusive version of the search space and pheromone intensities.
%
%
\subsection{Solving Phase II (Rescue): Signal routing and victim extraction}\label{subsec:SARinMaze:PhaseII.Extract}
In a Search and Rescue, once the trapped person, or the target is located, the MANet nodes reorganize themselves into a relay network to inform the base station of the target coordinates and call for help. In this phase, for computation purposes, the scattered MANet nodes can be treated like a graph \ensuremath{G}: a data structure consisting of two sets \ensuremath{(V, E)}, where \ensuremath{V} is the set of vertices and \ensuremath{E} is the set of edges connecting any two vertices in the graph.

A graph is usually represented as a diagram where vertices are symbolized as points and edges as lines connecting its end vertices \citep{west2001introduction}.

\begin{figure}[h]
\centering
  ~~~~~~~~~~~~~~~~\includegraphics[width=0.4\hsize]{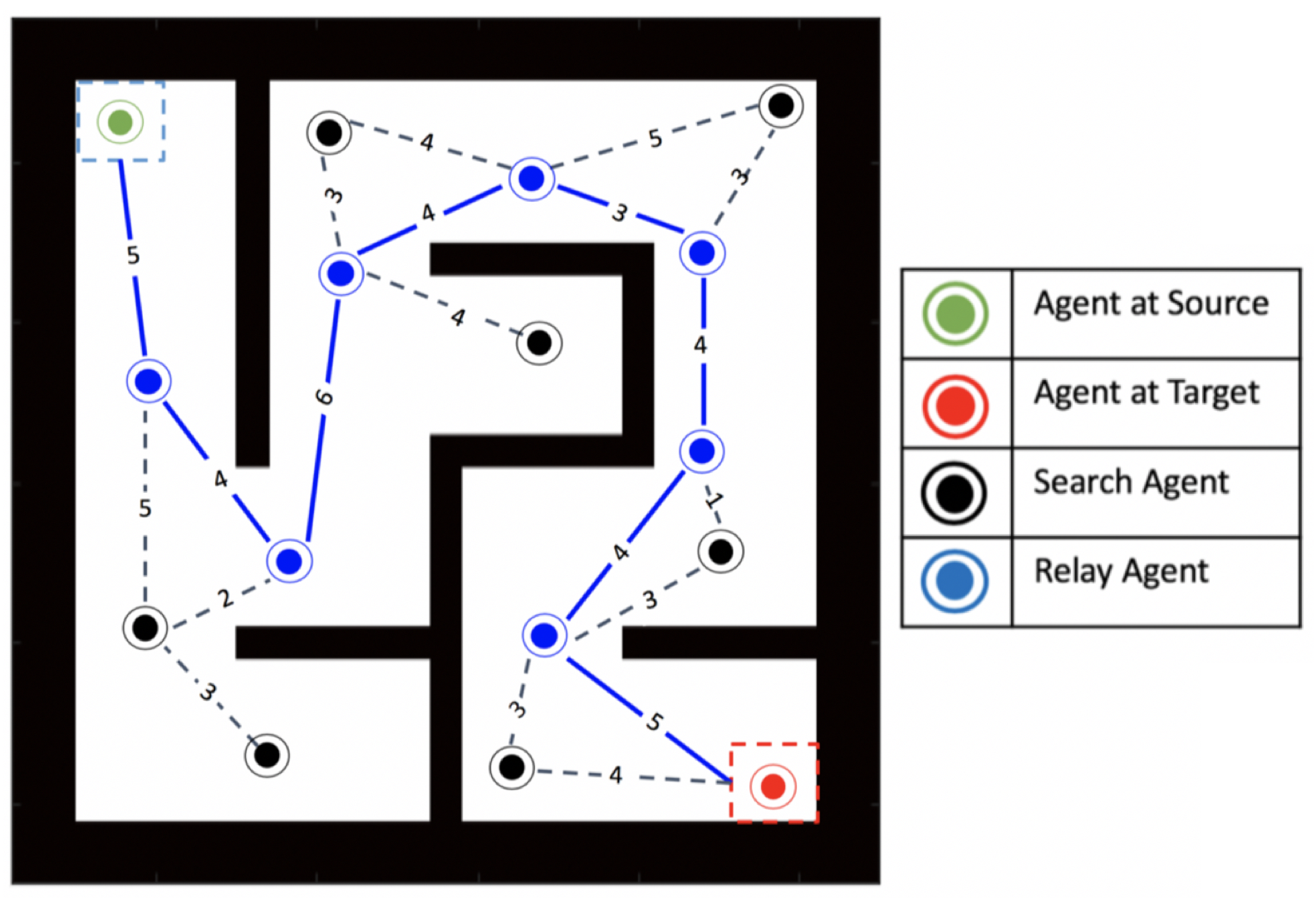}
\caption{The path cost calculations in the process of finding an optimal path back to the source. At each hop, the algorithm favors the link with the lowest cost associated to it.}
\label{fig:13}       
\end{figure}

Graphs have two main traversal approaches: Depth First Search (DFS) and Breadth First Search (BFS). DFS approach starts graph traversing at an initial node which is called a root and then goes down one path (branch) till the end of it, then it backtracks until the root and goes for another path to explore it. That is done until the graph is fully traversed. On the other hand, BFS starts at an initial root node then moves layer by layer in the graph where a node is explored first with its neighbors then explores node in the next layer. These layers are called depth-levels \citep{west2001introduction}.
%
%
\subsubsection{The Approach}\label{subsubsec:SARinMaze:PhaseII.Extract:Approach}
A graph is formed after exploration with agents as nodes and the edges are links between agents.
Links represent Line-of-Sight and Transmission Range.
From this graph and its adjacency matrix (the weights of the edges) we generate the \textit{Cost Matrix}.
The weights of the edges are evaluated based on the Euclidean distance between two agents (\ensuremath{u} and \ensuremath{v}) defined in the following equation:

\begin{equation}\label{eq:1}
Euclidean Distance(u,v)= \sqrt{ (x_u - x_v)^2+ (y_u - y_v)^2 }
\end{equation}

In an ideal scenario, on a successful conclusion of Phase I, the agents would be already interspersed in the maze (or search space) with  no big voids or obstacles between them and a path, back to the origin, can be easily found using the classic Dijkstra algorithm. However, the presence of obstacles or non-uniformity in the arrangement of the agents, in the form of large voids that disrupt communication links, a position based algorithm is proposed find a feasible complete path. The algorithm focuses on stopping the agents in the right locations to achieve a better spread of the agents in the environment.
%
%
%
\subsubsection{The Algorithm} 
To keep track of the diffusion of agents through the maze, we make use of a
an individual measure of depth. In a tree data structure, the depth of a node is simply the number of edges from the tree’s root to the node itself \citep{giannopoulou2009tree} and therefore easily calculated.
%



\begin{figure}[h]
  \includegraphics[width=0.95\hsize]{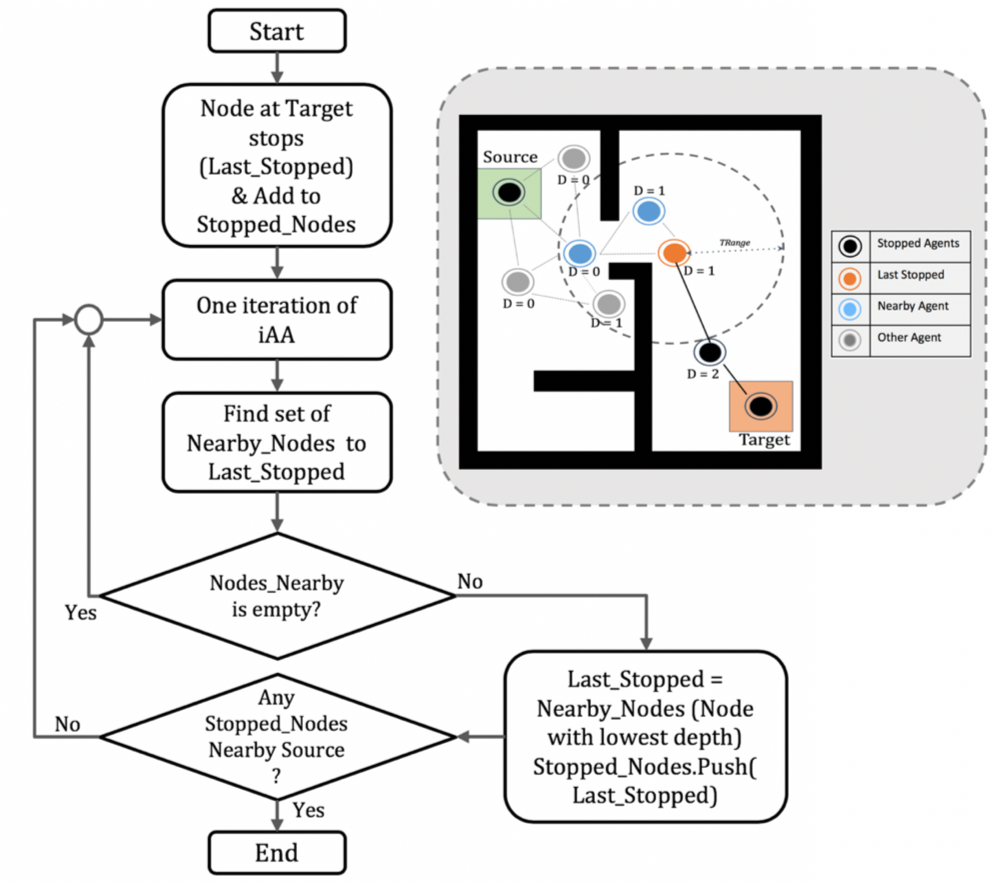}
\caption{Flowchart describing the working of the system in phase II (cf. Figure \ref{fig:two.phases.SAR}). The last stopped node considers only the nodes within its communication range as its neighbors in the calculations. if there are no nodes within its communication range, the iAA simulation is restarted to move the nodes until at least one neighbor appears.}
\label{fig:14}       
\end{figure}

\input{Algorithms/alg.iAA2}
In the maze-routing problem, the same concept can be applied where the tree’s root is the source and the depth there will have the value \ensuremath{0}. As the agents move away from the point of origin, some of them will lose direct access to the source either because of maximum range constraint or because of line-of-sight if walls break connections.
As a rule, agents will chose from all connected nodes that with the lowest depth, and then sets its own depth to this \ensuremath{(\text{depth} + 1)}.
The individual depth of agents that concurrently and iteratively apply this rule is the number of hops to the entry point (explicitly identifying the path).
\newpage
Figure \ref{fig:14} shows a flowchart of the working of the system in Phase II. The \textit{Last\ensuremath{\_}stopped} agent shown, is the one that was last chosen to be stopped by a previous stopped node and has to choose an agent from the nearby set of agents which abide by the following requirements:

\begin{itemize}
\item It is within the transmission range (\textit{TRange}) of the \textit{Last\ensuremath{\_}stopped} Agent.
\item There are no obstacles in between it and the \textit{Last\ensuremath{\_}stopped}.
\end{itemize}

In Figure \ref{fig:14} 
the first agent has depth of \ensuremath{0} (i.e., it can see the source). The other agent has depth of \ensuremath{1} (i.e., one hop away from the source). The next agent to be chosen is the one with the lowest depth (here: depth \ensuremath{0}). After the node is stopped
(within communication range and not blocked by obstacles)
the search ends and Dijkstra is used to find an optimal return path. 

The solution proposed is considered a semi-distributed model. At first the agents move in a distributed manner for the exploration part, then when the target is located the agents will broadcast a message that target is located back to source which will then find the shortest path in a centralized fashion.

If no path is constructed at first the agents will continue with the proposed improvement that is distributed. Agents will communicate when target is found and then the graph is constructed where source and target are included in it, if not, then the agent at target try to construct a path by searching for agents nearby until source is part of this path. Finally, the source can communicate to agents which agents form the optimal path. Figure \ref{fig:14} explains how the agents communicate when target is located.

%
%
\subsubsection{Routing}
\begin{figure}[ht]
  \includegraphics[width=\hsize]{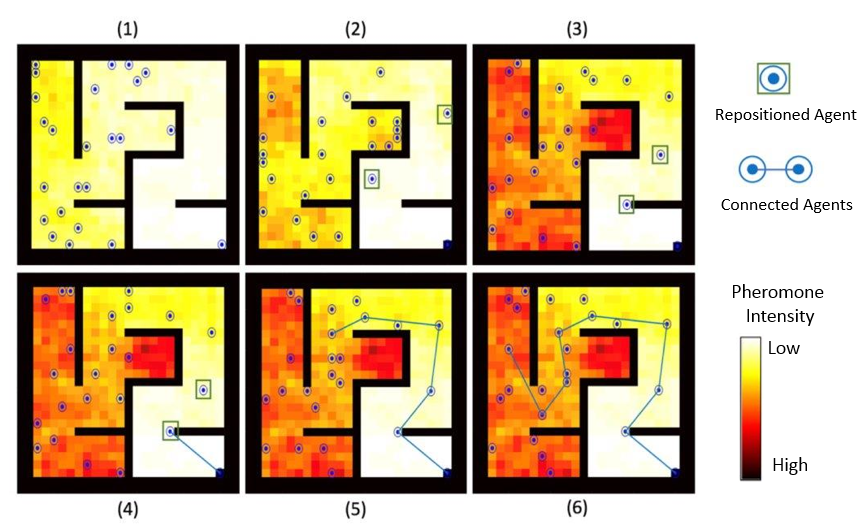}
\caption{A visual example showing the proposed approach: in (1) the target is located and the agent stops and starts looking for nearby agents, which (2) continue exploring. In (3) one agent is nearby, in (4) a link to that agent is created and it stops. This process continues (5) to other agents, using the least depth as guiding factor until (6) the process ends when the reached agent is near the source, causing the Dijkstra algorithm to terminate.}
\label{fig:15}
\end{figure}

For a simple simulation, we can assume the walls are impenetrable to the signal exchanged between the agents, and that the agent should have line of sight requirement to establish a connection to other agents. Figure \ref{fig:15} shows the scenario where a path initially is not constructed when the target is reached because there is no subgraph connecting the source to target.

In Figure \ref{fig:15} (1) the target is found (node circled in green); however, when calculating the distances which is the cost matrix and finding the possible connections, there is no path found between source and target nodes. The exploration continues using the iAA except for the node at target as it stops and tries to connect to nearby nodes (Figure \ref{fig:15} (2)). When a node is nearby the stopped node (Figure \ref{fig:15} (3)) it is able now to connect to it and it stops it.

The last stopped node identifies nearby nodes. Then, the model continues doing the same with every node connected in the sub-graph until one of the nodes is nearby the source (Figure \ref{fig:15} (6)). After reaching this state, Dijkstra re-runs,
to ensure path optimality.

\begin{figure}[h]
\centering
  \includegraphics[width=0.6\hsize]{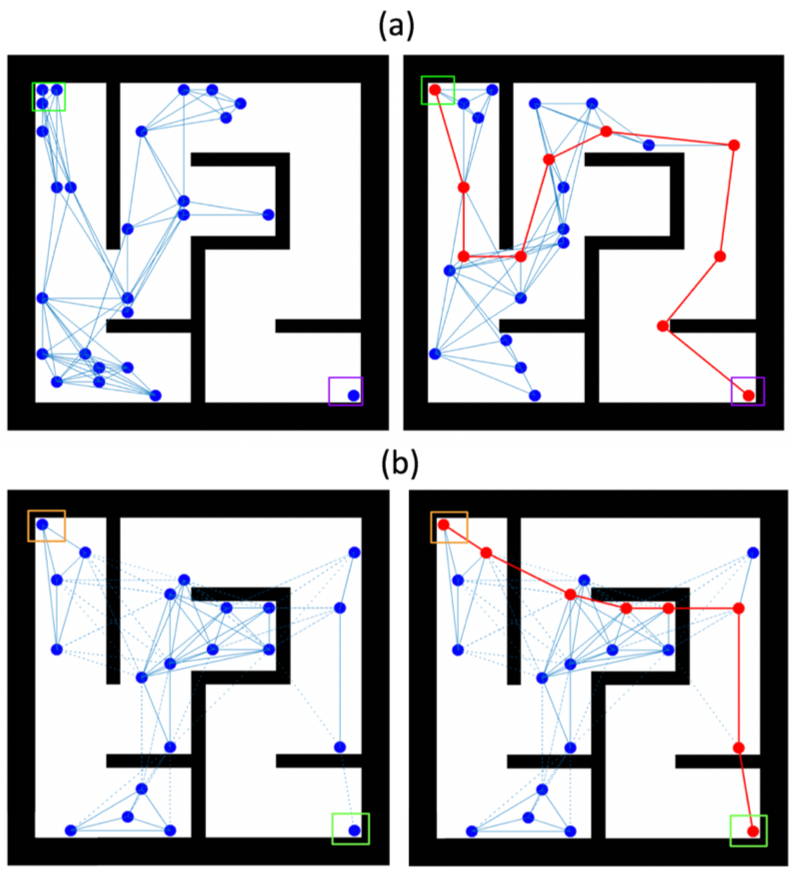}
\caption{Reverse route set up using the modified Dijkstra approach with implementation of both (a) impenetrable and (b) penetrable walls. The path length for the latter is much shorter than for the former. However, this is heavily dependent on the material of the walls in the search space, indicating it might influence the performance of our technique.}
\label{fig:16}
\end{figure}

Figure \ref{fig:16} (a) shows the situation of the agents before and after the routing performed. The agents marked in red are the part of the developed relay network. The remaining agents can be called back to the base station without affecting the performance of the relay.
\section{Materials and Methods}

%
%
\subsection{Modelling choices}

%
%
\subsubsection{Maze size and complexity}\label{subsubsec:MaM:Models:Mazes}
As stated above, we model a collection of mazes with different layouts, sizes and complexities.
To simplify the computation and implementation of the system, the mazes were discretized using the grid method similar in \cite{yang2011review}. The maze is decomposed into a grid, where the unit grid size is the same as the agent size, thereby making the step size equal to a unit grid.

\begin{figure}[h]
\centering
  \includegraphics[width=0.9\hsize]{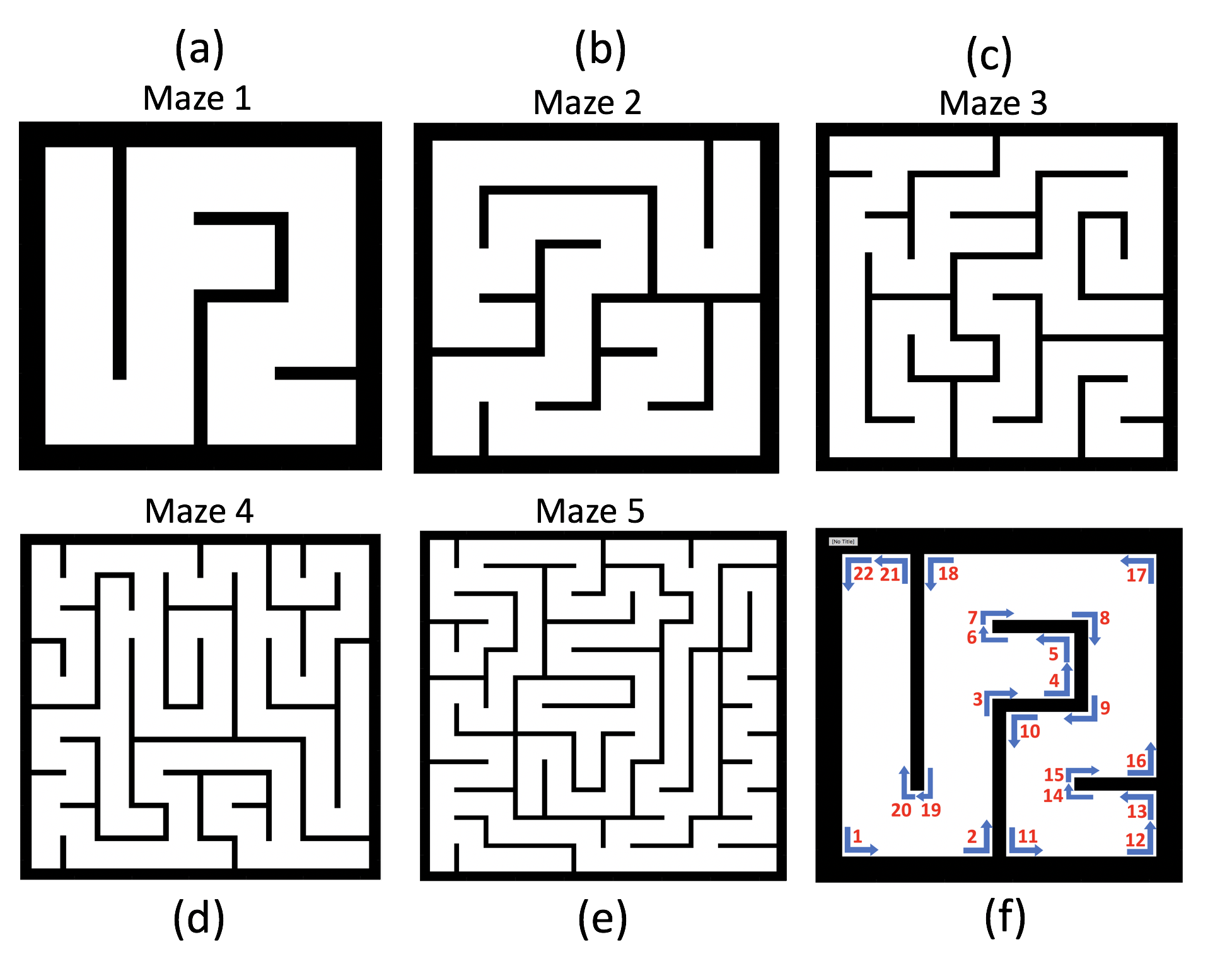}
\caption{(a)-(e) show the complexities of the 5 sample mazes (henceforth: M1-M5) referenced in this paper, cf. Table \ref{table:maze.complexity} for a numerical comparison. These are provided there as examples, in total 10 different layouts were generated for each complexity listed in Table \ref{table:maze.complexity} to further gauge the affect of randomness on the results. Maze (b) was previously used in Figure \ref{fig:2}. Panel (f) illustrates how we calculate maze complexity, using the example of maze (a): the complexity is the number of 90 degree turns an agent can make in the maze, cf. (f).}
\label{fig:different.mazes}
\end{figure}

This discretization helps visualize the system as a matrix which makes computation of signal reception, agent localization, and move generation easier. Naturally, future work\label{future:work:AgentsAndGridNotEqual} would address the situations where the size of the agent and the size of the grid unit are not equal.
Figure \ref{fig:different.mazes} shows
examples.

For each of the 5 types of mazes, 10 different layouts were generated randomly. Table \ref{table:maze.complexity} summarizes these maze sizes and provides a comparison of the respective complexities.
We define maze complexity as the number of 90 degree turns an agents can make.
Table \ref{table:maze.complexity} provides the values for our mazes.

\input{Tables/table.maze.complexity}

%
%
\subsubsection{Signal progression through obstacles}\label{subsubsec:MaM:Models:Routing}
To model indoor signal propagation, the ITU model, proposed by the International Telecommunication Union, described in \cite{Rappaport09}, was modified through the addition of the \ensuremath{(w \times c)} term to account for attenuation caused by the internal walls/obstacles to the following form:
\begin{equation} \label{eq:4}
PL[dB]=20log_{10}(f)+28 log_{10}(d)-28 + (w \times c)
\end{equation}
where \ensuremath{w = 4.4349} is the wall attenuation factor (for brick wall) \citep{wilson02propagation}, \ensuremath{c} is the wall count encountered, \ensuremath{f = 2400} MHz is the frequency channel of communication for standard Wi-Fi, and \textit {d} is the distance, in meters, from the beacon source \citep{Rappaport09}. Figure \ref{fig:5} shows the signal propagation
for different beacon placements. 

In real life situations, a radio wave is usually able to propagate through obstacles (depending on the thickness and material of the obstacle), but the signal strength significantly reduces due to a phenomenon called \textit{path loss} for which we use the ITU model 
to model an indoor beacon propagation. This path loss, can be in effect simulated by artificially lengthening a link across a wall. Accordingly, Equation \ref{eq:4} is rewritten to calculate the re-interpreted distance between the agents as:

\begin{equation}
Distance (m) = 10^{(\frac{PL + 28 - 20log_{10}(f)}{20})}\\
\end{equation}

This value is then fed into the \textit{Cost Matrix} (introduced in a simplified form in Section \ref{subsubsec:SARinMaze:PhaseII.Extract:Approach}). Due to this, the \textit{Cost Matrix} represents the new weight of the edge between 2 nodes and the Dijkstra algorithm, as we already described,  is ran to find the shortest path from the target back to the source with the possibility that the signal will penetrate the walls. 
A visual representation of how walls can affect signal strength was shown in Figure \ref{fig:5}; the impact of allowing sufficiently strong signals into the path even though the signal goes through a wall can be seen in Figure \ref{fig:16}, where the top version does not include signals that pass through walls, while the bottom does.

%
%
\subsection{Data collection}
Searching in an obstacle course such as a maze with a heuristic technique such as an Ant Algorithm is a process \textit{``rich in randomness''}, leading to excessively random search statistics (in terms of run time, number of steps, etc.). Therefore, in order to properly assess the performance of our implemented techniques, we carry out simulations on a collection of mazes with different layouts, sizes and complexities, each with 50 repetitions averaged out to realize the extent and appropriateness of the variations in the results. Figure \ref{fig:different.mazes} on page \pageref{fig:different.mazes} shows the 5 different maze complexities that were simulated, with Table \ref{table:maze.complexity} (on the same page) 
providing a comparison of the complexity values for these types. For each of these 5 types a total of 10 different layouts were generated to further gauge the affect of the randomness.

To study the impact of AA on the autonomous ants' led search in the maze, a simple ACO based algorithm along with a standard random search were implemented for comparison.

A set of experiments were set up to simulate AA and iAA algorithms with variable group sizes ranging between 100 and 600, with 100 unit increments.

\begin{figure}[h]
\centering
  \includegraphics[width=0.9\hsize]{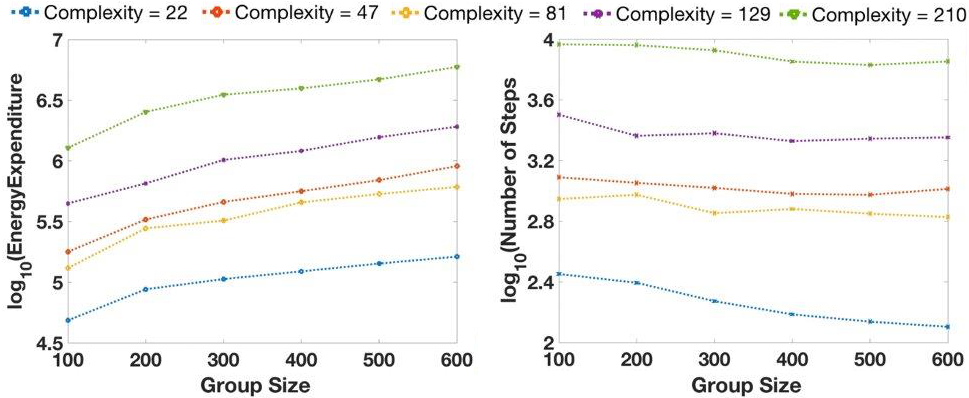}
\caption{Number of steps to solve the maze and the associated energy expenditure as a function of search group size used to solve the maze, for 5 different maze complexities.}
\label{fig:10}       
\end{figure}

Figure \ref{fig:10} (a) and (b) visualize the trend of the energy expenditure for 5 different maze complexities (measured, as shown in Figure \ref{fig:different.mazes}, in terms of 90 degree turns encountered) of 22, 47, 81, 129 and 210 (cf. Table \ref{table:maze.complexity}). There is a 30\% average increase in energy expenditure in solving the least and most complex mazes for the same group size. The decrease in number of steps needed to solve the maze is more obvious in the larger mazes, with an approximate 27\% decrease in the largest maze. The energy expenditure increases almost linearly with an increase in the number of agents in the search group.
\input{Tables/table.energy.expenditure}

We note that the entries in the Table \ref{table:energy.expenditure}, while commonsensical, are somewhat arbitrary and it is relatively easy to imagine a practical situation where a different cost applies to specific directions choices.  This, in turn, introduces some quantitative changes to the output in Figure 10.

%
%
\subsection{Performance measures}
    %
    %
%
%
\subsubsection{Benchmark comparison}
As a benchmark,
we compared the performance of our solution with another study that used ACO coupled with Fuzzy Logic and also presented results in terms of number of moves needed to solve the maze \cite{Mohit10}. For the purpose of benchmark comparison, we set up mazes to the other study with the same sizes of \ensuremath{8\times8} and \ensuremath{15\times15} units. See Section \ref{subsubsec:Results:PhaseI:Explore:Benchmark} for the results.
%
%
\subsubsection{Cumulative effort (steps)}
We equate the cumulative work of the swarm to
the number of steps taken by them. Agents act once per iteration but we only count this if it resulted in displacing the agent (i.e., we ignore the action of staying put).
The summation of the steps taken by all agents in all iterations leading to the discovery of the target is the measure of the cumulative effort of the team of agents.
%
%
\subsubsection{Estimated energy cost}
Similar to the above,
is the energy expenditure of the swarm, for example due to the limitations in search agents' battery lifetime. The agents' battery powers both its mechanical and communication functions, which means a faster depletion of the portable energy source and makes the issue crucial in the success of the operation. As elaborated below, the energy expenditure of each possible change to the ongoing linear forward motion (which itself is represented as ``one step forward'') is accounted for through a modification of the forward motion by a multiplicative factor, presented in Table \ref{table:energy.expenditure}. The cost for moving backward is highest, as we envision the backward motion as a complete halt followed by a reversal in direction, which would require the maximum energy.

%
%
\subsubsection{Assumptions and considerations}
Considering all agents expend comparable amount of energy for communication with each other, we approximate it as a constant for a particular group size and neglect it in energy calculations to simplify simulations at this stage of analysis. The assumption is that knowing a group size allows designers to add the energy needed for the communication as an additive constant.

The main variable component in energy consumption between agents of a group is the directed motion of the search agents through their consecutive steps. Continuing in the same direction of motion is always cheaper than introducing a displacement in direction. As our agents are limited to a four-directional motion (in addition to an option to hover), the only direction changes possible are a 90\ensuremath{^{\circ}}, to either side, or a 180\ensuremath{^{\circ}} turn, meaning a backward motion. Considering the different amounts of energy needed to slow down (control speed), and change direction in each motion possible, a motion based cost system was developed as shown in Table \ref{table:energy.expenditure}.
    %
    %
\section{Results and Discussion}

%
%
\subsection{Phase I: Maze exploration}
\begin{figure}[h]
\centering
  \includegraphics[width=0.9\hsize]{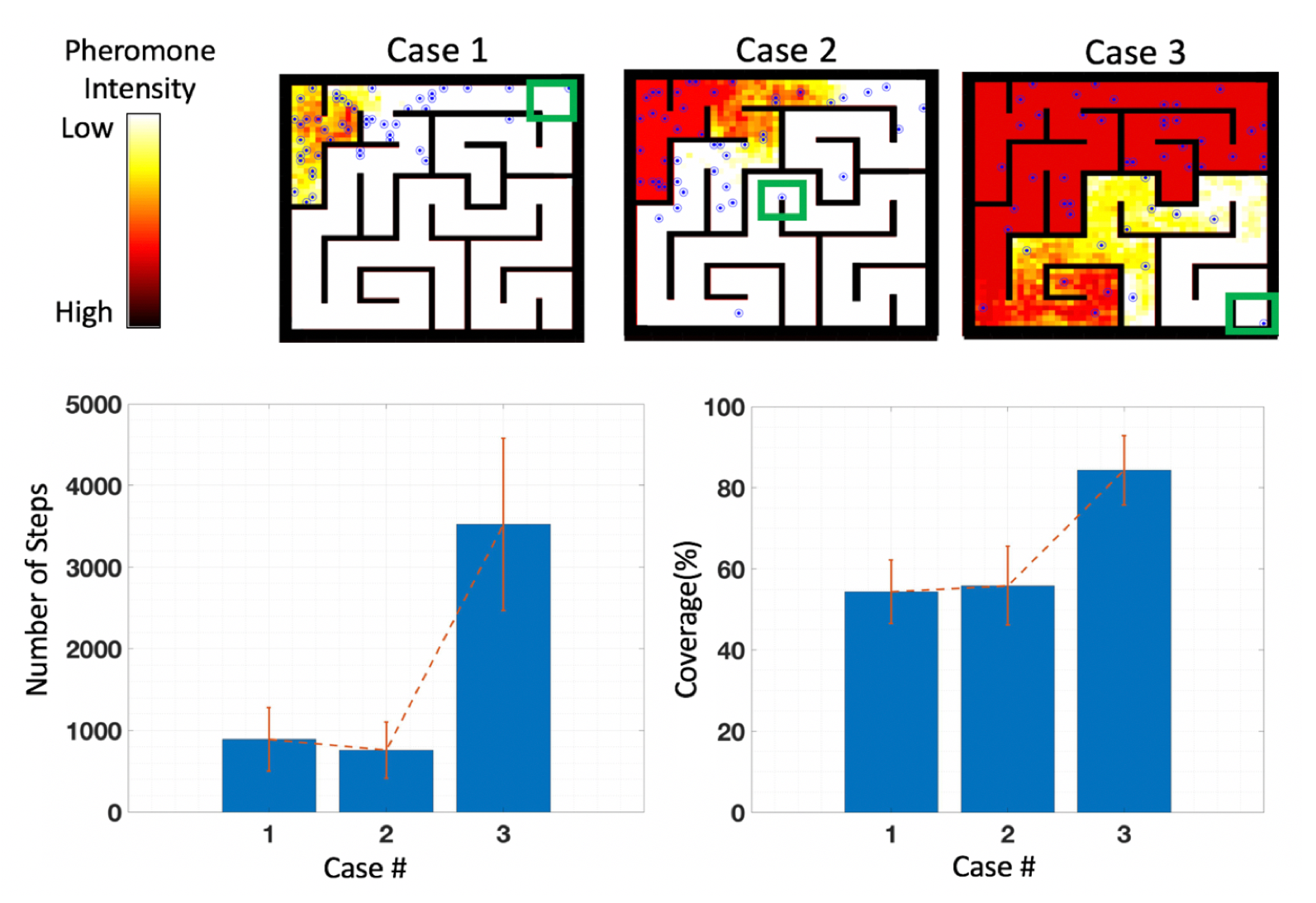}
\caption{Maze Coverage, measured the percentage of explored cells in the maze, as a by-product of the search process when tested with targets located at 3 different locations.}
\label{fig:12}
\end{figure}

Although maze exploration was not a key motive for the research, it is of interest to discuss how well the heuristic part of the iAA system covers the maze during the search, as the mapping ability could lead to other applications in the search process. As can be expected, the search results in almost a complete exploration (more than 80\%) of the search space, in a special case depicted by Case 3 in Figure \ref{fig:12}, when the target is located at the very end of the perfect maze. Exploration is much lower in cases where the target was located either close to the start or midway through the maze, but is still above 50\%, which implies that the iAA technique could also be used for that purpose.

%
%
\subsubsection{Benchmark comparison}\label{subsubsec:Results:PhaseI:Explore:Benchmark}
\begin{figure}[h]
\centering
\includegraphics[width=\hsize]{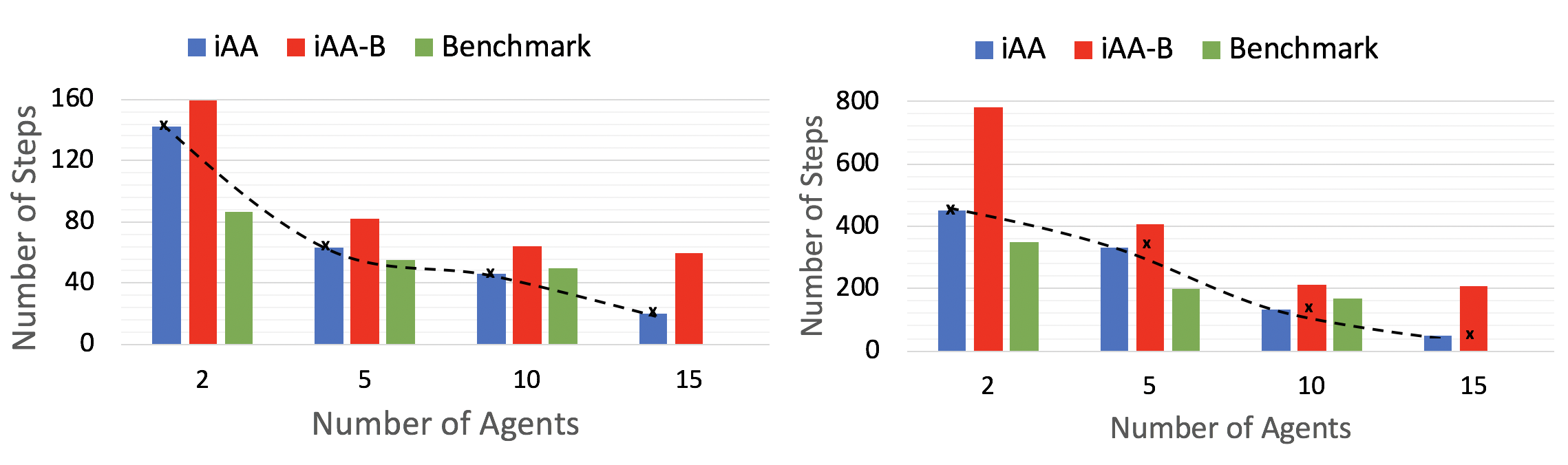}
\caption{While the benchmark outperforms our approaches (iAA and iAA-B) for a small group sizes, our technique iAA beats the benchmark \cite{Mohit10} as the swarm size increases. Furthermore, an extended study shows that the trend of improvement in iAA is even more favorable as the swarm size continues to increase. This change in performance can be attributed to the swarming effect which is better observed for a considerable search group size and doesn't show as much for a very small group (2 agents or so).}
\label{fig:11}
\end{figure}

The performance of our iAA model was poorer than the benchmarked technique for both maze sizes, when tested with search group sizes of 2 agents and 5 agents, which is very small size for a swarm, as can be seen in Figure \ref{fig:11}. We emphasize we chose such small groups to make a comparison with the closest relevant reference. However, our iAA algorithm was quicker to converge and needed fewer steps, when the group size was increased to just 10 agents. A likely reason for the better performance in our system is due to the poor exploration capability of small groups sizes, such as those with 2-5 agents.

For the maze sizes used, at least 7-10 agents would be necessary to promote quicker exploration that would lead to better results, as can be seen from the extrapolation of the trends in the results in Figure \ref{fig:11}. These small group sizes were tested only for comparison with the results presented in the benchmark. The slope of our solution, between different agent group sizes (measured by the number of agents in a group, cf. Figure \ref{fig:11}), is more favorable than the benchmark, as can be seen from Figure \ref{fig:11}. iAA also shows a qualitatively same behavior in a larger maze (right panel of Figure \ref{fig:11}) with quantitatively better performance with 10 or more agents (quantitatively better than in the case of the smaller maze). Therefore, the more the agents, and the bigger the maze, the more likely our proposed system is to outperform the benchmark study. Lastly, we point out that our iAA-B model performs less well, as could have been guessed from data in Figure 7 by extrapolating from large search group used in that study towards much smaller ones (in other words, the comparison of the performance between iAA and iAA-B holds across the large set of values of the number of search agents parameter).
\begin{figure}[h]
\centering
  \includegraphics[width=0.5\hsize]{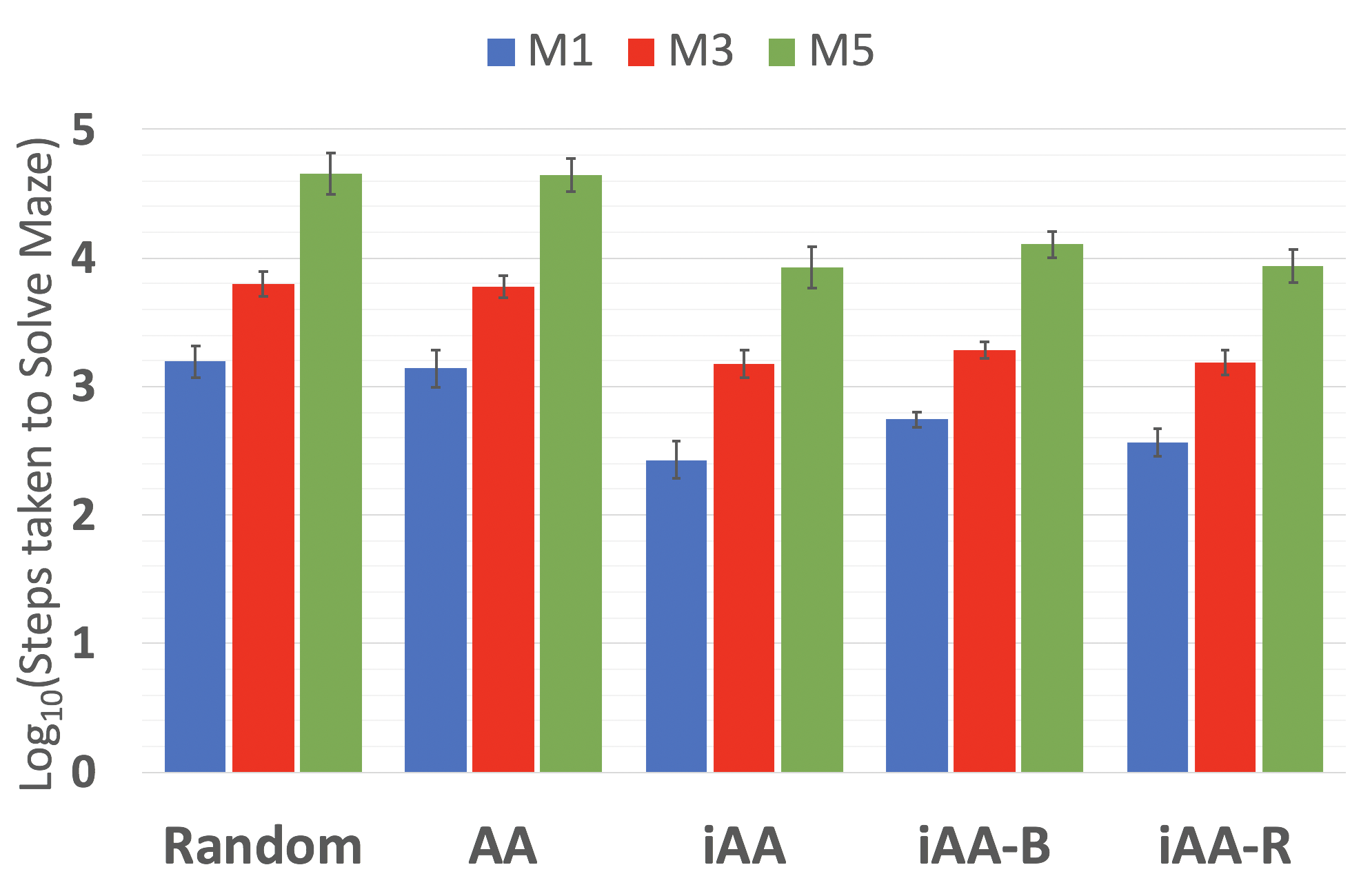}
\caption{Comparing the performance of the 4 AA-based models and a random movement solution in solving the 3 of the sample mazes (M1, M3 and M5), which are of different sizes and complexities, with 100 ants each.}
\label{fig:7}       
\end{figure}
%
%
\subsubsection{Cumulative effort (steps)}
Figure \ref{fig:7} (previous page) shows a comparison between the performances of the 5 models (when simulated with 100 ants each on the 3 mazes, with 30 repetitions each). The pure AA based model did not introduce much of an improvement, compared to a purely random solution. This lack of improvement can be attributed to the clustering effect of pheromone in pure AA that is likely limiting the exploration of the maze. iAA is the best performing algorithm among the 5, closely followed by the performance of iAA-R.

\begin{figure}[h]
\centering
  \includegraphics[width=6cm, height=6.5cm]{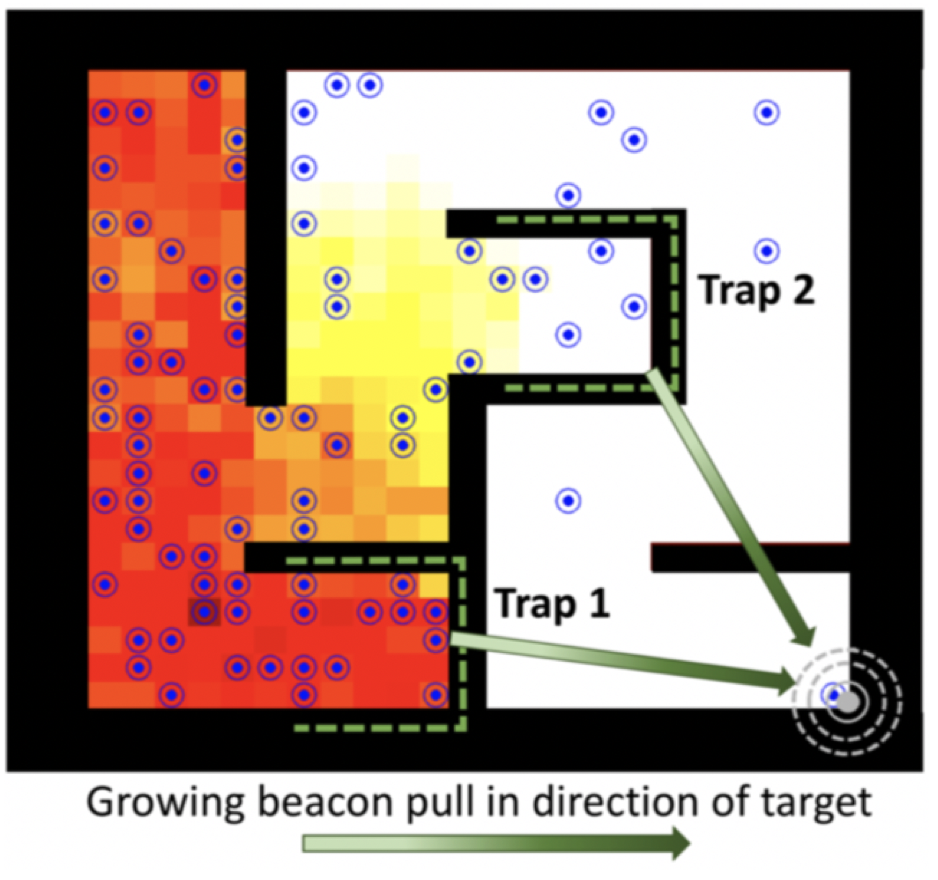}
\caption{Inverted AA with beacon initialization for pheromone intensities can lead agents into the traps, marked as green-dashed U-shaped walls in the figure, as they tend to blindly follow a positive pheromone gradient and are unable to go around these obstacles. }
\label{fig:8}
\end{figure}

Contrary to expectations, iAA-B did not positively add to the performance of the iAA algorithm with its beacon initialization that was supposed to better guide the ants to the target. This is due to a trapping effect noticed in the simulation, as illustrated in Figure \ref{fig:8}, where ants get trapped in nooks of the maze while being pulled towards the target.

\begin{figure}[h]
  \includegraphics[width=0.5\hsize]{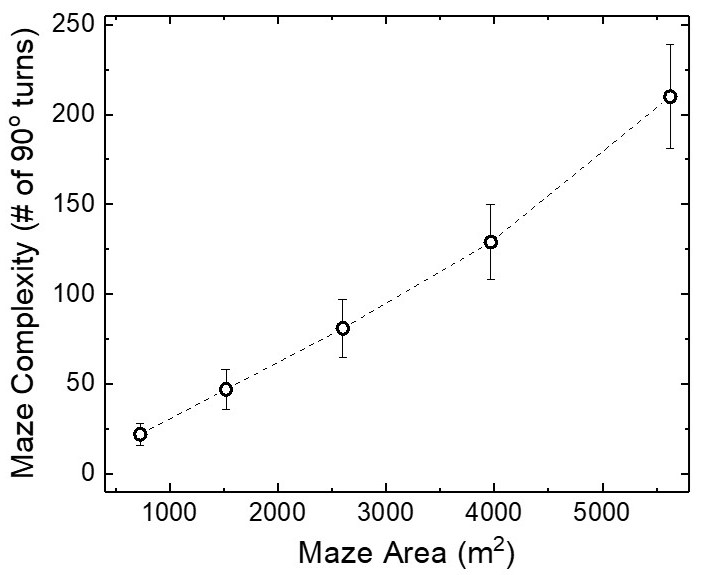}
  \includegraphics[width=0.5\hsize]{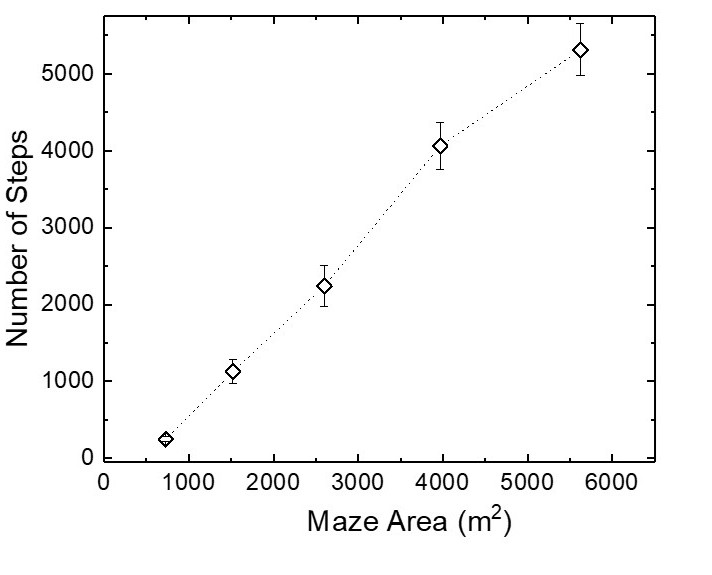}
\caption{Performance of iAA algorithm as a function of variable maze size and maze complexity (cf. Table \ref{table:maze.complexity}), with a constant search group of 100 agents.}
\label{fig:9}
\end{figure}

As shown in Figure\ref{fig:9}(a), the growth in maze complexity is weakly supra-linear with the increase in maze size (cf. Table \ref{table:maze.complexity} on page \pageref{table:maze.complexity}). The performance of the iAA strategy, measured by the number of necessary steps, in Figure \ref{fig:9}(b), with a fixed search group size, initially grows very fast with the increase in maze area, but appears to approach a saturation-like behavior, above a certain maze size. Based on this and insights from related numerical experiments, this point of near saturation determines the optimal maze size and complexity problem that can be solved by a search group of 100 agents.

The weakly supra-linear dependence of the maze complexity on the maze size is to be expected, given that the number of ``topological'' choices (e.g. how to get around the maze), grows faster than the size itself. This can be checked through simple counting, by developing versions of Figure \ref{fig:different.mazes}(f) while allowing the maze size to increase. The suggestion of the existence of the optimal size of the search agent group (Figure \ref{fig:9}(b)) for a fixed maze size (while averaged overall a number of configurations at a fixed size), implies the presence of the non-linear dependence of the \textit{Number of Steps} on the \textit{Number of Agents}.

Based on our numerical experiments
it appears that the Search Task (Phase I) could have ``too few'' or ``too many'' agents, which is understandable in the context of the ant agent repellent action.
%
%
\subsubsection{Estimated energy cost}
%
%
\subsection{Phase II: Signal routing and victim extraction}
\begin{figure}[h]
  \includegraphics[width=\hsize]{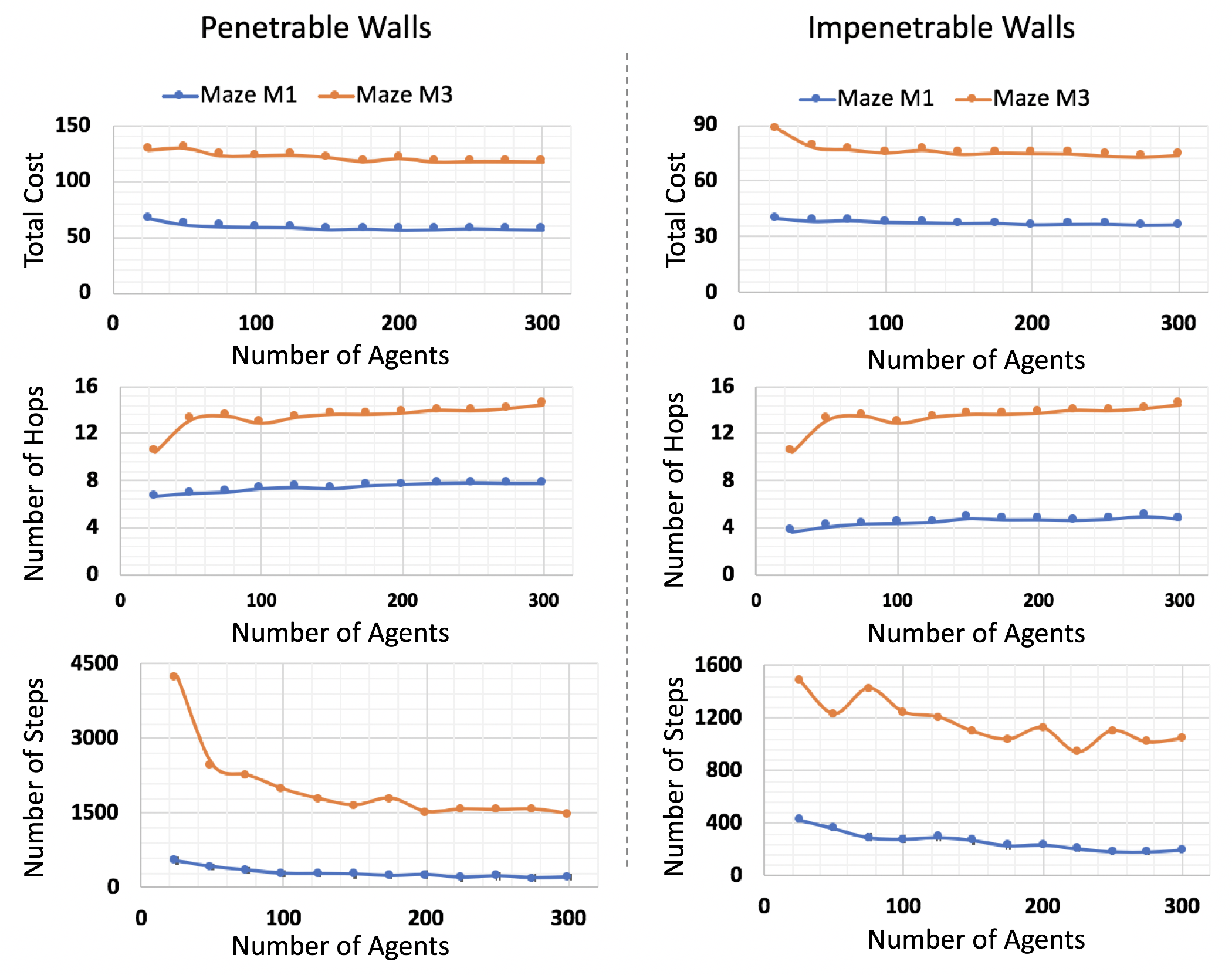}
\caption{Performance analysis of Phase II in terms of number of steps (or iterations of the algorithm) taken to solve the maze, number of hops in the final relay network formed, and the total cost of the relay path found (measured in distance), as tested for 2 maze complexities, M1 and M3.}
\label{fig:17}       
\end{figure}

To study the impact of changing the number of agents on the final solution in Dijkstra-based Phase II (the ``rescue'' part in SAR), the three performance metrics variables were generated - the total cost, the number of hops, and the number of steps, in all cases as function of the number of agents. These metrics are shown in Figure \ref{fig:17}.  Here, number of steps implies the number of iterations of the algorithm execution to find a solution, while number of hops refers the number of logical hops in the relay network formed.

For the steps count, one can intuitively assume that a small number of agents requires more steps to achieve the task then it slowly decreases as the number of agents increases and converges to a certain value. The reason is that is fewer agents will need more time to form the shortest path as the probability for it to have more voids and go through the proposed search with stopping algorithm is higher. The number of agents is not strongly affecting the time of reaching the target as it depends on what path agents will follow. For the total cost, it decreases slightly when increasing the number of agents.

By observing the simulation examples figure, one can deduce that the path looks almost the same; however, for the larger agents' population the dependence is more smooth. Furthermore, the total cost when penetrable walls are considered is much less than impenetrable. The number of steps is also lower for penetrable walls but not by the same factor as the cost difference.

We have analyzed the performance for two different maze types, M1 and M2. As we have seen earlier in the paper, both size (area) and complexity of the maze play role in the performance (for Phase I).

Figure \ref{fig:17} shows that both parameters are relevant in the performance of Dijkstra-based reverse routing. 
The dependence is neither simply linear nor smooth, as evidenced by non-monotonous behavior of the number of Steps and Hops for maze M3 (it being more complex) than for the maze M1.
    %
    %
    %
    %
\section{Conclusion and Future work}

%
%
\subsection{Conclusion}

\begin{figure}[h]
  \includegraphics[width=\hsize]{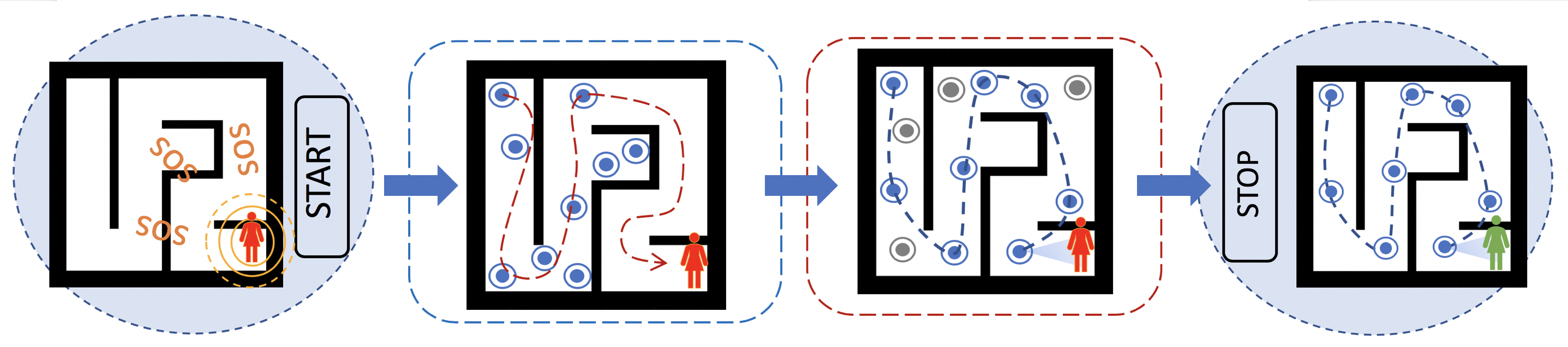}
\caption{A summary view of the SAR process in two phases. The search starts with a victim being trapped in the maze-like environment, broadcasting an SOS signal. In Phase-I the search agents explore the environment in search of the SOS signal, using iAA algorithm. After locating the source of SOS signal, the agents switch to Phase-II, and find the shortest routing path to communicate information about the victim's location to the base station. The remaining unused agents (marked in gray) can return to the base station. }
\label{fig:18}       
\end{figure}

A summary of the SAR performed in 2 phases, is summarized pictorially in Figure \ref{fig:18}. The search starts with a victim being trapped in the maze-like environment, broadcasting an SOS signal. In Phase I, the search agents pro-actively explore the environment in search of the victim, using the iAA algorithm. Once the victim is spotted, the agents switch to the Phase II, and find the shortest routing path to communicate information about the victim's location to the base station, relying on specific implementation of Dijkstra algorithm and the results of the Phase-I search.
Three different forms of the Ant Algorithm were explored and tested, to be able to devise the optimal search method.

The influence of the maze size and complexity, as well as the number of the search agents are studied, together with the study of energy resources' dependence on problem and environment parameters.
In ongoing efforts, we study the dynamic (time-varied) injection of the agents as a possible path towards additional performance improvement. We are in contact with the agencies that may use this approach in their SAR training.
%
%
\subsection{Future Work}
In Section \ref{future:work:AgentsAndGridNotEqual} we stated that we  simplified the computation and implementation of the system by discretizing a grid method similar to the one used in \cite{yang2011review}. Naturally, future work would address the situations where the number of agents (swarm size) and the size of the grid unit are not equal.

Furthermore, the simulation is a 2D representation of the world. Extending this to a 3D model (both for the signal propagation through obstacles as well as with the layout and structure of the building in mind) will be subject to application driven need to do so. The presented implementation and performance evaluation suffices to argue for the approach \textit{in theory}, a real-world implementation will have to address a number of engineering issues as well as account for the 3D nature of the world. As we acknowledged before, there is a very active community working on all sorts of approaches related to e.g.,
path finding for multiple devices \citep{ICAPS20paper26}, collision free movement of a swarm \citep{10.5555/3398761.3399020}, movement in formation \citep{10.5555/3398761.3398848} or through narrow spaces or bottlenecks \citep{ICAPS20paper49}. Whether it be in 2D or 3D, we fully expect that any deployed  solution using our approach 
will consist of a merger of multiple techniques, tailored to the specific problem at hand \citep{9300245}. As we discuss in \citep{SwarmIsMoreThanSumOfDrones.2021}, applying nature inspired solutions \textit{for the sake of it}, while popular, is problematic. Future work will consider the practical aspects of implementing and deploying our solution for a swarm of UAVs and reflect on how our approach can best be combined with both, nature-inspired \citep{Berlingereabd8668} as well not nature-inspired, approaches to further optimize performance.

Finally, in future work we would like to allocate some time to investigating the complexity of the mazes further. Preliminary investigations where we compared optimal path lengths in randomly created mazes yielded results that informed the work presented in this article, however, the results themselves were omitted, mainly for the sake of brevity. A real-world study of mazes and operational environments for SAR swarms is required for work targeting the actual deployment of a drone swarm.
\begin{acknowledgements}
We acknowledge the UAE ICTFund grant ``Bio-inspired Self-organizing Networks'', and we thank Prof. Sami Muhaidat (KUST) for his advice about path loss models. We thank Prof. Patrick Grosfils (Universite Libre de Bruxelles) and Prof. R. Mizouni (KUST) on encouraging exchanges.
\end{acknowledgements}
\bibliographystyle{spbasic_updated}      
\bibliography{ms}
\end{document}

%% file: Tables/table.litstudy.heuristics.tex
\begin{table}[h]
\centering
\begin{tabular}{lccc}
\hline\noalign{\smallskip}
Ref & \textbf{\textsc{LP}} & \textbf{\textsc{HT}} & Algorithm Used  \\
\noalign{\smallskip}\hline\noalign{\smallskip}
\citep{wang2017trajectory}
    &  & $\checkmark$ & Improved ACO\\
\citep{malone2017hybrid}
    & $\checkmark$ &  & Artificial Potential Field Method\\
\citep{atten2016uav}
    & $\checkmark$ & $\checkmark$ & Multi Pheromones for tracking targets\\
\citep{Cao16}
    & $\checkmark$ & $\checkmark$ &Improved ACO Heuristic Function\\
\citep{Krentz15}
    &  & $\checkmark$ & Simple ACO\\
\citep{fossum14repellent}
    & $\checkmark$ & $\checkmark$ & Repellent Pheromone for coverage\\
\citep{deepak2014advance}
    & $\checkmark$ & $\checkmark$ & Advanced PSO\\
\citep{WangR16}
    &  & $\checkmark$ & Genetic Algorithm with ACO\\
\citep{Aurangzeb13}
    &  & $\checkmark$ & Hybrid ACO w. Random- + RL based-Search\\
\citep{buniyamin11}
    & $\checkmark$ &  & Point Bug Algorithm\\
\citep{wang2011application}
    & $\checkmark$ &  & Dijkstra Algorithm\\
\citep{Mohit10}
    & $\checkmark$ & $\checkmark$ & Fuzzy Logic with Counter ACO\\
\citep{gong2009robot}
    &  & $\checkmark$ & PSO in partially known environments\\
\citep{sauter2005performance}
    & $\checkmark$ & $\checkmark$ & Combination of multiple pheromones\\
\noalign{\smallskip}\hline
\end{tabular}
\caption{Research in the field of deterministic and heuristic path planning strategies in maze like environments including Ant Colony Optimization (ACO), Particle Swarm
optimization (PSO), and others. \textbf{\textsc{LP}} stands for \textit{Local Planning}, \textbf{\textsc{HT}} stands for \textit{Heuristic Technique}.}
\label{table:litstudy.heuristics}
\end{table}

%% file: Tables/table.litstudy.dijkstra.tex
\begin{table}[h]
\centering
\begin{tabular}{clcc}
\hline\noalign{\smallskip}
&Reference & Cost Function & Alg.  \\
\noalign{\smallskip}\hline\noalign{\smallskip}
$\checkmark$ & \cite{ma2018path}
    & Distance, Gas Concentration  & \textbf{D}, \textbf{ACO} \\
& \cite{fawzy2017balanced}
    & Energy & \textbf{D}\\
&\cite{gupta2016applying}
    & Distance  & \textbf{D}\\
$\checkmark$ & \cite{fadzli2015robotic}
    & Energy, Difficulty, Distance & \textbf{D}\\
$\checkmark$ & \cite{liu2012floyd}
    & Distance & \textbf{FD}\\
$\checkmark$ & \cite{raihybrid2011}
    & Distance  & \textbf{D}, \textbf{ACO}\\
$\checkmark$ & \cite{yazici2006heuristic}
    & Distance & \textbf{D} + others \\
\noalign{\smallskip}\hline
\end{tabular}
\caption{Research using deterministic and heuristic path planning strategies; ($\checkmark$) indicates maze-like environments with obstacles. The used algorithms are \textbf{D}ijsktra, \textbf{F}loyd-\textbf{D}ijkstra, \textbf{A}nt \textbf{C}olony \textbf{O}ptimization. \textit{others} refers to a combination of sweep and savings.}
\label{table:litstudy.dijkstra}
\end{table}

%% file: Algorithms/alg.iAA.tex
\begin{algorithm}
initialization\;
possible moves = [stay, right, left, forward, backward]\;
\While{target not found}
{
\For{each ant}
{
Generate list of all possible next states\;
Acquire pheromone information of all next states\;
Roulette Wheel $\leftarrow$ generate probabilities of moving to each next state\;
Spin Roulette Wheel to pick next state\;
Update current position and pheromone levels\;
}
}
\caption{The basic AA Algorithm. 
The AA, iAA, iAA-B, and iAA-R models are based on the same algorithm; they differ mainly in the equation used for probability distribution generation.}
\label{alg:iAA}
\end{algorithm}

%% file: Algorithms/alg.iAA2.tex
\begin{algorithm}[h]
$Last\_Stopped$ = First agents to arrive at target\;
$Stopped\_Agents$ = [$Last\_Stopped$]\;
$Possible\_Moves$ = [right, left, forward, backward]\;

\While{Source not within $Stooped\_Agents$}
{
	\For{each agent not in $Stopped\_Agents$}
    {
    	Generate list of all possible next states\;
        Acquire pheromone information of all next states\;
        Roulette Wheel $\leftarrow$ generate probabilities of moving to each next state\;
        Spin Roulette Wheel to pick next state\;
        Update current position and pheromone levels\;
        $Agents\_Nearby$ = [All agents that could connect to $Last\_Stopped$]\;
        $Last\_Stopped$ = Min(depths of $Agents\_Nearby$)\;
        $Stopped\_Agents.push(Last\_Stopped)$;
    }
}
end function;

\caption{Determining the shortest path to the victim.}
\label{}
\end{algorithm}

%% file: Tables/table.maze.complexity.tex
\begin{table}[h]
\centering
\caption{Dimensions and average complexities of the 5 maze types used in simulations.}
\label{table:maze.complexity}
\begin{tabular}{lcc}
\hline\noalign{\smallskip}
 & Dimensions & Complexity  \\
\noalign{\smallskip}\hline\noalign{\smallskip}
Maze 1 (M1) & 27 $\times$ 27 & 22\\
Maze 2 (M2) & 39 $\times$ 39 & 47\\
Maze 3 (M3) & 51 $\times$ 51 & 81\\
Maze 4 (M4) & 63 $\times$ 63 & 129\\
Maze 5 (M5) & 75 $\times$ 75 & 210\\
\noalign{\smallskip}\hline
\end{tabular}
\end{table}

%% file: Tables/table.energy.expenditure.tex
\begin{table}[h]
\centering
\caption{Cost estimates for all different possible directions / actions of movement.}
\label{table:energy.expenditure}

\begin{tabular}{cc}
\hline\noalign{\smallskip}
  Direction & Cost  \\
\noalign{\smallskip}\hline\noalign{\smallskip}
    Hover & 0.5 \\
    Forward & 1 \\
    90\ensuremath{^{\circ}} turn  & 1.5 \\
    Backward & 2 \\
\noalign{\smallskip}\hline
\end{tabular}
\end{table}